\documentclass[sn-basic]{sn-jnl}

\usepackage{graphicx}%
\usepackage{multirow}%
\usepackage{amsmath,amssymb,amsfonts}%
\usepackage{amsthm}%
\usepackage{mathrsfs}%
\usepackage[title]{appendix}%
\usepackage{xcolor}%
\usepackage{textcomp}%
\usepackage{manyfoot}%
\usepackage{booktabs}%
\usepackage{algorithm}%
\usepackage{algorithmicx}%
\usepackage{algpseudocode}%
\usepackage{listings}%
\usepackage{aas_macros}

\theoremstyle{thmstyleone}%
\theoremstyle{thmstyletwo}%

\theoremstyle{thmstylethree}%

\begin{document}

\title[Article Title]{Accretion Geometry of Black Hole X-ray Binaries: Insights from X-ray Observations}


\author*[1]{\fnm{Honghui} \sur{Liu}}\email{honghui.liu@uni-tuebingen.de}



\affil*[1]{\orgdiv{Institut f\"ur Astronomie und Astrophysik}, \orgname{Eberhard-Karls Universit\"at T\"ubingen}, \city{T\"ubingen}, \postcode{D-72076}, \country{Germany}}




\abstract{The accretion-ejection activities of black holes play a vital role in shaping the Universe. 
Bright and recurrent black hole X-ray binaries are ideal objects for studying accretion physics 
across a wide range of accretion rates, providing insights into the understanding of their supermassive 
counterparts. This short review summarizes X-ray techniques capable of measuring accretion geometry, 
our current understanding, and open questions. In particular, X-ray spectroscopic studies indicate that 
the accretion disk can extend close to the innermost stable circular orbit in the bright hard state. Some 
hints of disk-corona-jet connections are also discussed.}

\keywords{X-ray binaries, Black holes, accretion}



\maketitle

\section{Introduction}\label{sec1}



Black hole (BH) X-ray binaries (XRBs) are systems in which a stellar-mass black hole, with typical masses ranging from several to tens of 
solar masses ($M_{\odot}$), is accreting matter from its companion star (see Figure~\ref{disk_corona}). The accreted matter 
forms a disk-like accretion flow around the central BH. This process is highly efficient at converting gravitational energy into radiation, producing strong electromagnetic emission. In the case of accretion onto stellar-mass BHs, the emission is predominantly in the X-ray band. These X-ray photons are particularly valuable, as they originate from 
the innermost regions of the accretion flow near the BH. As a result, the X-ray signal serves as a powerful tool for studying 
BHs and the nearby strong gravity region \citep{DeRosa2019SCPMA..6229504D, Reynolds2021, Bambi2021}.

The first confirmed BH XRB, Cygnus X-1 \citep{Murdin1971Natur.233..110M}, was discovered during the early years of X-ray astronomy by instruments onboard sounding rockets \citep{Bowyer1965Sci...147..394B, Bolton1972Natur.235..271B}. Since then, space-based X-ray all-sky monitors (e.g., RXTE, MAXI) have continued to discover new BH XRB candidates. As of today, around 80--100 candidates have been identified in the Galaxy, of which approximately 20 have been dynamically confirmed to host stellar-mass black holes.\footnote{Two useful catalogs include: (1) XRBcat: \url{http://astro.uni-tuebingen.de/~xrbcat/index.html}; (2) BlackCAT: \url{https://www.astro.puc.cl/BlackCAT/index.php}.} \citep[][]{Tetarenko2016ApJS..222...15T, Corral-Santana2016, Neumann2023A&A...677A.134N, Avakyan2023A&A...675A.199A}.

Early X-ray satellites revealed complex spectral and timing behaviors, which led to the initial definition of spectral states in BH XRBs \citep{Tanaka1995xrbi.nasa..126T}. Later missions with high energy and timing resolution, such as \textit{RXTE} (2--60 keV), \textit{XMM-Newton} (0.15--12 keV), \textit{Chandra} (0.1--10 keV), \textit{INTEGRAL} (3 keV -- 10 MeV), \textit{Swift} (0.2--10 keV for XRT and 15--150 keV for BAT), \textit{Suzaku} (0.2--12 keV for XIS and 10--600 keV for HXD), \textit{NuSTAR} (3--79 keV), \textit{NICER} (0.2--12 keV), \textit{Insight}-HXMT (1--250 keV), and \textit{AstroSat} (0.3--100 keV), have provided a more comprehensive and unified understanding of these states, based on their combined spectral and timing properties \citep[see][for detailed reviews]{Remillard2006, McClintock2006csxs.book..157M, Done2007, Belloni2010, Belloni2016ASSL..440...61B}.


In addition to the high-quality data provided by these missions, there has been significant progress over the past 15 years in modeling the X-ray signals from BH XRBs, particularly regarding the spectral and timing signatures of the relativistic reflection component \citep[e.g.,][and see Sec.~\ref{spec}]{Garcia2010, Ingram2019MNRAS.488..324I, Bambi2021} and the interpretation of quasi-periodic oscillations \citep[e.g.,][and see Sec.~\ref{timing}]{Ingram2019NewAR..8501524I}. These improvements collectively contribute to a better understanding of the accretion flow geometry, which is essential for addressing more fundamental questions such as measuring black hole spins \citep[e.g.,][]{Reynolds2021, Liu2023ApJ...950....5L, Draghis2024ApJ...969...40D} and potentially even testing
general relativity \citep[e.g.,][]{Bambi2017, Liu2018JCAP...08..044L, Liu2019, Bambi2024PPN....55.1420B}. 

More recently, the launch of the Imaging X-ray Polarimetry Explorer (IXPE) in 2021 opened a new observational window via X-ray polarimetry. Its findings are already challenging our pre-polarimetry understanding of accretion geometry and dynamics (see Section~\ref{acc_geo}). Gravitational wave (GW) detections by LIGO, Virgo, and KAGRA have also provided insights into populations of black
holes found in BH-BH/BH-NS binaries, some of which may also be
stellar-remnant black holes \citep{Abbott2021PhRvX..11b1053A, Abbott2021ApJ...913L...7A,Abbott2023PhRvX..13a1048A}. Interestingly, the black holes detected through GWs appear to be more massive \citep[e.g.,][]{Abbott2023PhRvX..13a1048A, Liotine2023ApJ...946....4L} and to spin more slowly \citep[e.g.,][]{Fishbach2022ApJ...929L..26F, Connors2024FrASS..1092682C} than those found in XRB systems, indicating different BH populations or evolutionary pathways.

In this review, we summarize our current understanding of the accretion geometry in BH XRBs and highlight open questions that remain to be addressed. In Section~\ref{states}, we describe the main spectral components and states of BH XRBs. Section~\ref{techniques} outlines the X-ray techniques used to study accretion geometries. In Section~\ref{acc_geo}, we summarize the inferred accretion geometry based on current X-ray observations.

\section{X-ray emission and spectral states}
\label{states}

\begin{figure}[h]
    \centering
    \includegraphics[width=0.98\linewidth]{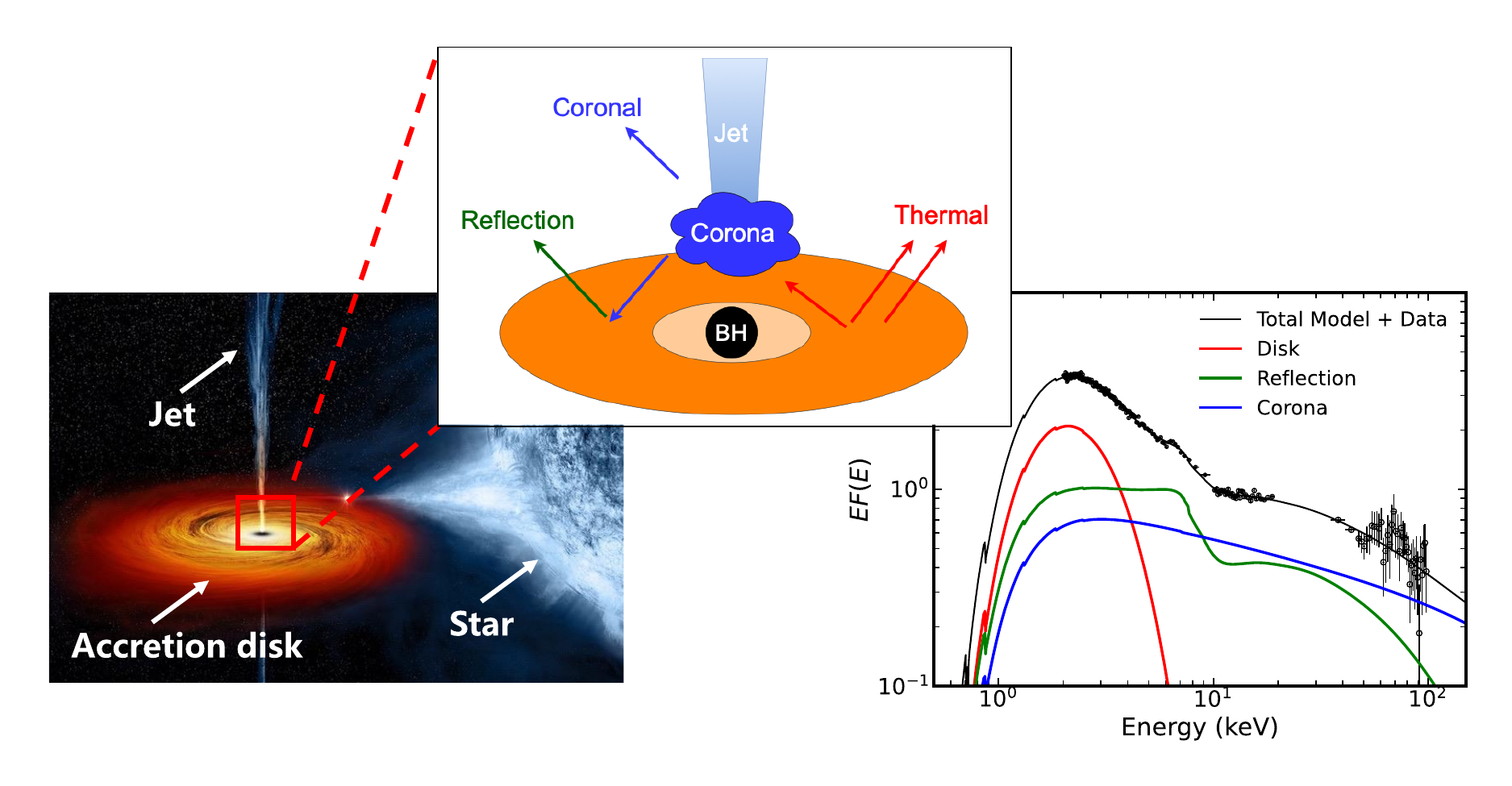}
    \caption{(Left) Artist's impression of a black hole X-ray binary system (courtesy of JPL/NASA). (Middle) A zoom-in sketch of the inner accretion flow near a BH. Arrows with different colors represent emission 
    components: (red) the disk thermal
    emission; (blue) the coronal emission; (green) the reflected emission of the disk. (Right) The Insight--HXMT 
    observation of the low mass BH XRB GX 339--4 in the intermediate state on
    2021 March 27. The red, blue and green lines show spectral components corresponding to the middle
    panel. The sum is shown as the black solid line. Figure adapted from \cite{Liu2024thesis}.}
    \label{disk_corona}
\end{figure}

The X-ray spectrum of BH XRBs typically consists of several components, as illustrated in Figure~\ref{disk_corona} in the 
framework of the disk-corona model. The central BH is surrounded by an optically thick, geometrically thin 
accretion disk \citep{Shakura1973, Novikov1973blho.conf..343N}. The accretion disk emits blackbody emission locally, 
and the spectrum of the whole disk is a multi-temperature blackbody \citep[e.g., disk thermal emission,][]{Mitsuda1984}. 
The blackbody temperature generally increases for radii closer to the black
hole, and can reach 1--2~keV at the innermost stable 
circular orbit (ISCO) radius for BH XRBs when they are accreting at appreciable
fractions of their Eddington limits.. The disk blackbody component typically peaks in 
soft X-ray band (e.g., at few keV or less, Fig.~\ref{disk_corona}).
Near the BH, there exists a cloud of hot plasma (with a temperature of $\sim$~100~keV), commonly referred to as 
the corona \citep{Thorne1975ApJ...195L.101T, Shapiro1976ApJ...204..187S}. The corona can 
Compton up-scatter seed photons from the disk, or from internal processes within the corona itself 
\citep{Malzac2009MNRAS.392..570M, Poutanen2009ApJ...690L..97P}, producing a non-thermal power-law-like 
emission that extends into the hard X-ray band. While the basic structure of the accretion disk is 
relatively well understood, the geometry and formation mechanism of the corona remain less certain 
\citep[see][for recent reviews]{Liu2022iSci...25j3544L, Bu2023arXiv231020637B}. 


The coronal emission can illuminate the optically-thick
accretion disk and be reprocessed to produce the \textit{reflection} 
component \citep{George1991,Garcia2010}. In the rest-frame of the accretion disk, the reflection spectrum is characterized 
by narrow fluorescent emission lines (the strongest one is typically the iron K${\alpha}$ line at 6.4--7 keV) and a Compton reflection hump that peaks around 20--30 keV. The observed reflection spectrum is broadened 
by relativistic effects (e.g., Doppler shifts and gravitational redshift, \citealt{Fabian1989}) and carries valuable information about the accretion system (see details in Sec.~\ref{spec}).

\begin{figure}[h]
    \centering
    \includegraphics[width=0.95\linewidth]{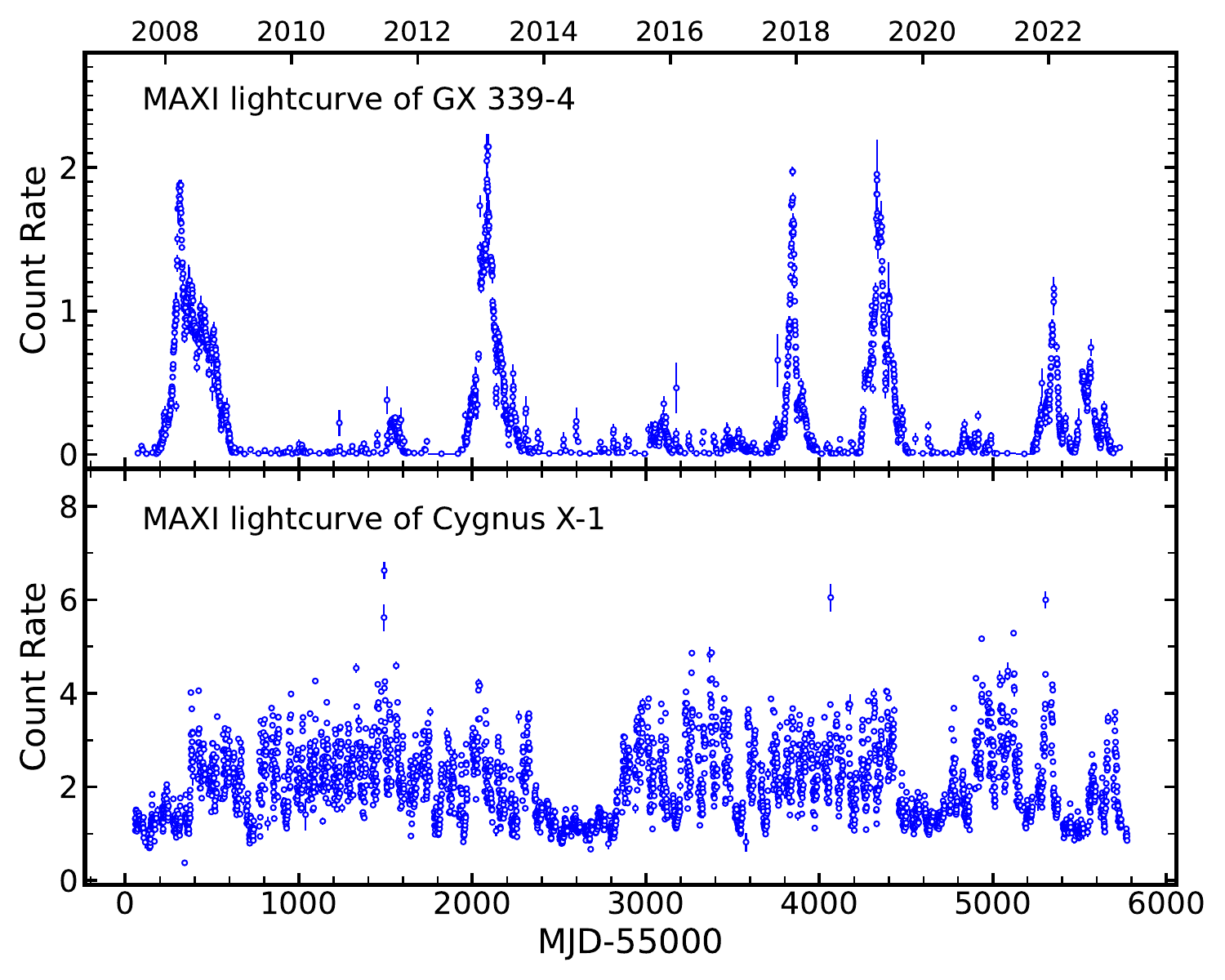}
    \caption{Lightcurves of a transient (GX~339-4) and a persistent (Cygnus~X-1) BH XRBs by the 
    Monitor of All-sky X-ray Image (MAXI) in the 2--20~keV band.}
    \label{lc}
\end{figure}

BH XRBs can generally be divided into persistent and transient sources, with the latter comprising the majority \citep{Corral-Santana2016}. Transient BH XRBs spend most of their time in the \textit{quiescent} state, characterized by 
low X-ray luminosity (e.g., $L_{\rm X}<10^{33}$~erg~s$^{-1}$) and a 
hard, non-thermal X-ray spectrum \citep{Kong2002ApJ...570..277K, Corbel2006ApJ...636..971C, 
McClintock2006csxs.book..157M, Reynolds2014MNRAS.441.3656R}. 
Figure~\ref{lc} shows typical lightcurves of a transient BH XRB (GX~339-4) and a persistent source (Cygnus X-1). Transient sources can undergo outbursts triggered by instabilities in the outer disk 
\citep{Lasota2001NewAR..45..449L, Coriat2012MNRAS.424.1991C} during which their X-ray luminosity 
can increase by several orders of magnitude (reaching $L_{\rm X}=10^{37}-10^{39}$~erg~s$^{-1}$). Persistent sources remain continuously bright and never enter quiescence.

The outburst can also be traced using the hardness-intensity diagram (HID, \citealt{Belloni2005, Belloni2010}). 
In this diagram, the hardness can be simply defined as the ratio of count rates in a hard X-ray band to those in a soft X-ray
band. It broadly indicates the spectral shape of the source. The intensity can be the total count rate of the detector, 
which is related to to the source luminosity or mass accretion rate. There is no strict definition of the hard/soft X-ray bands or
the intensity. Alternative diagrams, such as the rms-intensity diagram \citep{Munoz-Darias2011MNRAS.410..679M}, can 
also be used to track the outburst using information from both the source luminosity and timing variability. It is even possible to 
trace the outburst using timing properties only, such as the ``hue angle'' defined on the power spectral 
density \citep{Heil2015MNRAS.448.3339H}. The hue angle is defined on a power-color (PC) diagram. Two power colors are first defined:
(1) PC1: the variance in the 0.25--2.0 Hz band divided by the variance in the 0.0039--0.031 Hz band.
(2) PC2: the variance in the 0.031--0.25 Hz band divided by the variance in the 2.0--16.0 Hz band.
Observations have shown that during an outburst, the source traces an elliptical path on a PC2 versus PC1 diagram, centered at (PC1, PC2) = (4.5, 0.45) (see Figure 1 of \citealt{Heil2015MNRAS.448.3339H}). The hue angle is then defined as the clockwise angle from a reference axis at 45$^{\circ}$ to the \textit{x} and \textit{y}-axes (see Figure 2 of \citealt{Heil2015MNRAS.448.3339H}). The ranges of the hue angle corresponding to different spectral states are provided in Table 2 of \citealt{Heil2015MNRAS.448.3339H}.

\begin{figure}[h]
    \centering
    \includegraphics[width=0.98\linewidth]{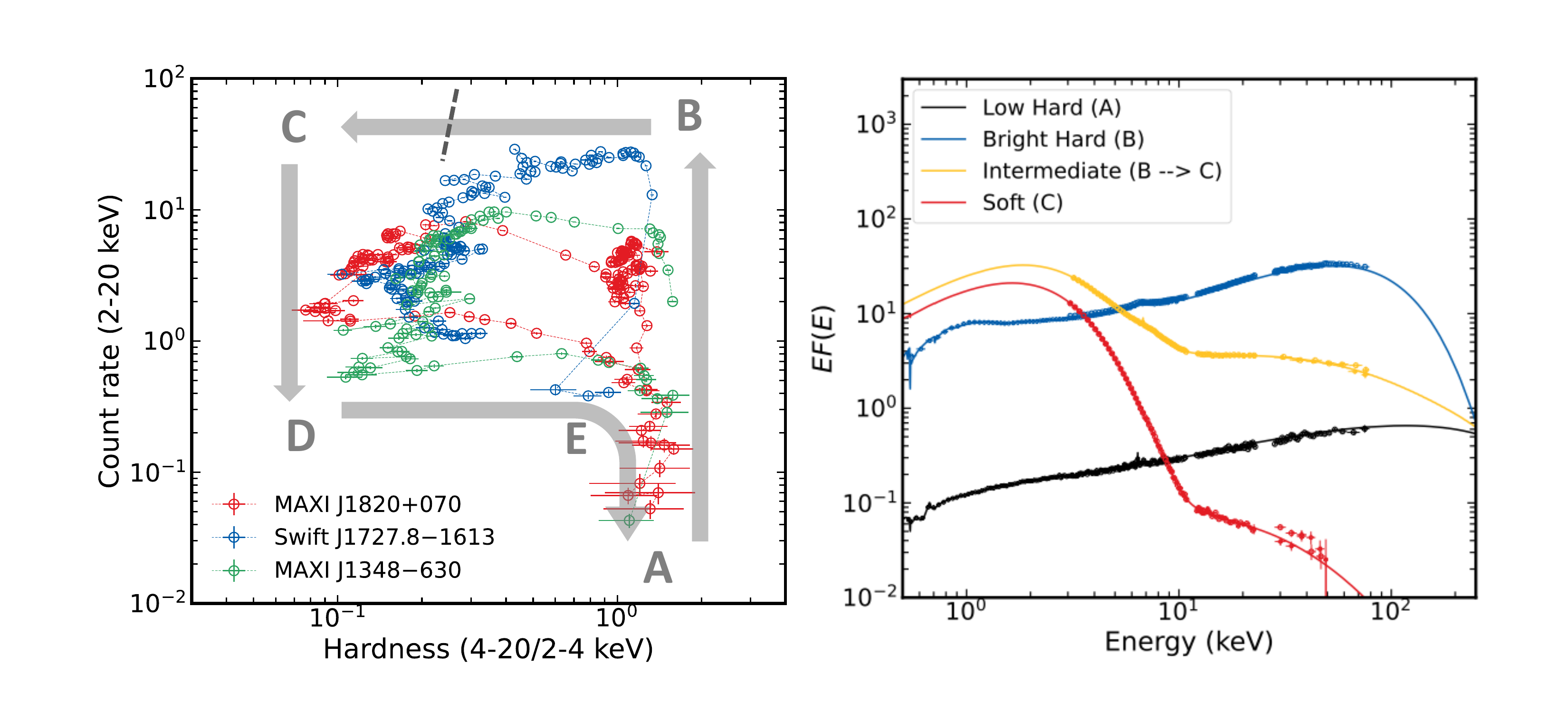}
    \caption{(Left) Hardness-intensity diagrams of three recent outbursts from transient BH XRBs 
    plotted using MAXI data. The dashed line
    denotes the launch of ballistic jets. Figure adapted from \cite{Liu2024thesis}. (Right) X-ray spectra of 
    different states from \textit{NuSTAR} and \textit{NICER} observations of MAXI~J1820+070.}
    \label{hid_spec}
\end{figure}

All of these different methods for characterising the time-evolving X-ray properties
of a system reveal how a source transitions between distinct \textit{states} during 
an outburst \citep[e.g.,][]{Fender2004,Homan2005,Remillard2006}. In Figure~\ref{hid_spec}, the outbursts of three sources 
are shown to follow a similar pattern on the HID. As 
the source emerges from quiescence, it first enters the lower-right corner of the HID (region A), indicating a low 
luminosity and a hard X-ray spectrum. This is known as the \textit{low hard} state during which the X-ray spectrum is 
dominated by a power-law-like component from the corona (see the right panel of Figure~\ref{hid_spec}), and the 
radio emisison indicates the presence
of a steady compact jet \citep{Corbel2003A&A...400.1007C, Kylafis2012A&A...538A...5K}. 
As the accretion rate increases, the source moves upwards on the HID into the \textit{bright 
hard} state (region B). During this phase, the source luminosity rises (as shown in Figure~\ref{hid_spec}), 
while the spectral shape remains largely unchanged and continues to be dominated by the coronal component. 

Then the source makes a leftward turn on the HID and transitions quickly to the 
\textit{soft state} (region C), passing through the short-lived hard intermediate and soft intermediate states. The transition luminosity 
varies from source to source and may even differ between outbursts of the same source. The state transition 
is manifested as an increasing contribution from the disk thermal component and a steepening of the coronal component in 
the spectrum. During this transition, a highly relativistic, episodic, 
and ballistic jet is often launched \citep[e.g.,][]{Kording2008Sci...320.1318K, Bright2020NatAs...4..697B}, 
and the steady compact jet is then quenched \citep[e.g.,][]{Russell2019ApJ...883..198R}. 
This roughly corresponds to the boundary between the hard and soft intermediate states (jet line in Figure~\ref{hid_spec}, 
\citealt{Fender2009MNRAS.396.1370F, Miller-Jones2012MNRAS.421..468M, Homan2020ApJ...891L..29H}).

In the soft state, thermal emission from the disk 
completely dominates the spectrum, while the coronal emission is diminished and the jet disappears. After spending 
most of the outburst time (weeks to months) in the soft state with steadily decreasing luminosity, the source 
transitions back to the low hard state (from region D to A) at a 
luminosity lower than that of the hard-to-soft transition. This ``hysteresis effect'' causes a ``q''-shaped track on the HID. 
Since the intensity roughly corresponds to the mass accretion rate, the hysteresis implies that mass 
accretion rate alone does not determine the spectral state and accretion
flow properties 
\citep[e.g., ][]{Meyer-Hofmeister2005A&A...432..181M, Liu2005A&A...442..555L, 
Petrucci2008MNRAS.385L..88P, Begelman2014ApJ...782L..18B}. It should also be noted that some outbursts do not go through 
the whole q-diagram and finish without reaching the soft state \citep[`failed' outbursts,][]{Tetarenko2016ApJS..222...15T, Alabarta2021MNRAS.507.5507A, Wang2022MNRAS.512.4541W}.
Moreover, some peculiar individual sources do not exhibit a canonical q-diagram during their outbursts, 
though states can still be defined based on their spectral and timing properties \citep[e.g.,][]{Belloni2000A&A...355..271B, Wijnands2000ApJ...528L..93W,  Wang2024ApJ...963...14W, Fan2024arXiv241207621F}.

Although persistent sources do not exhibit this kind of quiescence-outburst cycle, they still show strong 
variability in their lightcurves (see Figure~\ref{lc}). This variability can also be traced on the HID, 
where similar transitions between spectrally hard and soft states are observed 
\citep[e.g.,][]{Smith2007ApJ...669.1138S, Hirsch2020A&A...636A..51H, Jiang2024Galax..12...80J, Konig2024A&A...687A.284K}.

A common interpretation for the low hard state is that the accretion disk is truncated at a large 
radius \citep{Tomsick2009}, within which lies an optically thin, hot accretion flow that may act as the corona, 
producing strong power-law emission \citep{Narayan1994ApJ...428L..13N, Done2007}. In the high soft state, 
the accretion disk is thought to extend down to the innermost stable circular orbit (ISCO) \citep{Steiner2010}. 
However, significant debate remains regarding how the disk and corona evolve from the low hard to the 
high soft state, particularly through the bright hard and intermediate states. A key question is 
whether the disk remains truncated in the bright hard state, which has important implications 
for understanding the mechanism of state transitions. 

\section{X-ray techniques to measure accretion geometry}
\label{techniques}

In this section, we summarize the X-ray techniques that can be used to measure/infer the geometry of the disk-corona system.

\subsection{X-ray spectroscopy}
\label{spec}

X-ray spectroscopy involves the analysis of the X-ray spectra from sources, as shown in the 
right panels of Figure~\ref{disk_corona} and 
Figure~\ref{hid_spec}. 

\subsubsection{Spectroscopy of the Corona}

For coronal geometries, a variety of models have been proposed in the literature, as summarized in Figure 1 of \cite{Poutanen2018A&A...614A..79P}.
The basic geometry of the corona can be constrained from the shape of its spectrum.
In the quiescent state, when the X-ray luminosity is very low (e.g., $L_{\rm X}/L_{\rm Edd}<10^{-4}$), the photon index ($\Gamma$) saturates at approximately 2.1 \citep{Sobolewska2011MNRAS.417..280S, Plotkin2013, Reynolds2014MNRAS.441.3656R}.
This is likely because the spectrum is dominated by jet emission, such as that from inverse Comptonization or synchrotron emission from a non-thermal particle distribution \citep{Yuan2005ApJ...629..408Y}.
In the low hard state (region A of Figure~\ref{hid_spec}), the coronal spectrum is hard (i.e., the photon index $\Gamma\approx 1.6-1.8$) and its temperature is high ($\sim 100$~keV). This is only possible in a ``photon-starved'' geometry, where the cooling power provided by seed photons (e.g. from the disk) is not too high. Otherwise, the corona would be cooled down and its spectrum would become soft. This constraint requires the ratio of the heating power in the corona ($L_{\rm h}$) to the power of the seed photons illuminating it ($L_{\rm s}$) to be: $L_{\rm h}/L_{\rm s} \gg 1$.
In this case, the sandwich-corona (or slab-corona) geometry \citep{Haardt1991ApJ...380L..51H, Haardt1993ApJ...413..507H}, in which the corona fully covers the underlying cold disk, is ruled out. This is because half of the coronal emission is reprocessed by the disk and emerges as soft seed photons. This reprocessing forces $L_{\rm h}/L_{\rm s} \sim 1$ and leads to a soft spectrum ($\Gamma > 2$), even if all the accretion power is dissipated in the corona \citep{Stern1995ApJ...449L..13S, Poutanen1996ApJ...470..249P, Malzac2001MNRAS.326..417M}. Furthermore, intrinsic dissipation within the disk itself produces additional seed photons, resulting in even softer spectra. 

There are alternative scenarios that meet the $L_{\rm h}/L_{\rm s} \gg 1$ requirement:
\begin{itemize}
\item The truncated disk scenario, in which the standard geometrically thin, optically thick disk is truncated at a large radius. Inside this radius, a hot, geometrically thick, and optically thin accretion flow is present and provides
the high-energy `coronal' emission \citep{Esin1997}. This hot flow is generally believed to be an advection-dominated accretion flow (ADAF) \citep{Ichimaru1977ApJ...214..840I, Narayan1994ApJ...428L..13N}. In this geometry, the hot flow intercepts only a small fraction of the soft photons from the outer disk.
\item The patchy corona scenario, in which the corona consists of localized active regions on the surface of the disk \citep{Haardt1994ApJ...432L..95H, Stern1995ApJ...449L..13S}. Here, although half of the coronal emission is reprocessed by the disk, only a fraction of those reprocessed photons re-enter the corona as seed photons.
\item The outflowing corona scenario, in which the corona moves away from the disk with a relativistic bulk velocity \citep{Beloborodov1999ApJ...510L.123B, Malzac2001MNRAS.326..417M}. In this case, beaming reduces the X-ray emission directed toward the disk, resulting in fewer reprocessed soft photons. 
\end{itemize}

Among the three scenarios, the ADAF solution is only valid for low accretion rates (up to $L/L_{\rm Edd} \sim 0.01$ for a viscosity parameter $\alpha=0.1$; \citealt{Esin1997}). However, observations suggest the luminous hard state (Region B) can reach a much higher accretion rate ($L/L_{\rm Edd} \sim 0.2$; \citealt{Gierlinski2003MNRAS.342.1083G, Koljonen2016MNRAS.460..942K, Liu2023ApJ...950....5L}). The patchy-corona model naturally predicts comparable luminosities for the direct coronal emission and the reflection component, a prediction which contradicts observations in
the fainter stages of the hard state \citep[e.g.,][]{CadolleBel2007ApJ...659..549C, Nowak2011ApJ...728...13N, Furst2015, Wang2018, Wang2020ApJ...899...44W} but is consistent with data from the luminous hard state and intermediate states \citep[e.g.,][]{Miller2002ApJ...570L..69M, Miller2004ApJ...606L.131M, Xu2018maxij1535, Liu2023ApJ...950....5L}. Ultimately, spectral fitting alone is often insufficient to distinguish between different coronal geometries \citep[e.g.,][]{Brenneman2014ApJ...781...83B} or to obtain detailed quantitative constraints on the corona.

\subsubsection{Spectroscopy of the Disc}

For the disk thermal emission component, the disk temperature increases as the disk moves closer 
to the black hole. In the meantime, relativistic effects (e.g., Doppler boosting and 
gravitational redshift) also become stronger because of the faster orbital velocities. 
Therefore, the disk thermal component can potentially be used to measure the inner disk radius. Practically, this can be done 
using the simple Newtonian model \texttt{diskbb} \citep{Mitsuda1984} or more appropriate relativistic models such as
\texttt{kerrd}\footnote{\url{https://heasarc.gsfc.nasa.gov/xanadu/xspec/manual/node190.html}} \citep{Ebisawa2003ApJ...597..780E} 
and \texttt{nkbb}\footnote{\url{https://github.com/ABHModels/nkbb}} \citep{Zhou2019PhRvD..99j4031Z}. This method is 
most effective when the disk thermal component contributes significantly to the spectrum, such as during 
the intermediate and soft states. This technique is often referred to as \textit{continuum fitting}.
Using the continuum-fitting method alone to constrain the inner disk radius (or BH spin) requires prior knowledge of the BH mass, distance, and disk inclination angle\footnote{It has been shown via simulations that high-quality X-ray spectra allow the continuum-fitting method to simultaneously constrain the black hole spin and the inclination of the disk. This requires a well measured thermal disk emission across a wide energy range \citep{Parker2019MNRAS.484.1202P}.} \citep{Zhang1997ApJ...482L.155Z, McClintock2014SSRv..183..295M}.
This requirement limits the application of the method to sources where this information is known. One alternative is to implement continuum fitting together with the reflection method (see below). The reflection method can provide complementary constraints on the inner disk radius and disk inclination. Consequently, the priors on BH mass and distance can be relaxed, allowing these parameters to be measured using only X-ray data \citep[e.g.,][]{Parker2016, Zdziarski2025ApJ...981L..15Z, Das2025arXiv250909481D}. Moreover, due to scattering in the disk atmosphere, the local emerging spectrum at a given disk radius often deviates from a true blackbody. This effect is typically modeled with a modified or diluted blackbody spectrum \citep{Shimura1995ApJ...445..780S}. The local specific flux is given by $F_{\nu} = \pi B_{\nu}(f_{\rm col}T_{\rm eff}) / f_{\rm col}^4$, where $f_{\rm col}$ is the color correction factor, $T_{\rm eff}$ is the local effective temperature, and $B_{\nu}$ is the Planck function. The color correction factor is often assumed to be constant, i.e., independent of disk radius and accretion rate, in spectral fitting. Such an assumption is likely appropriate within a single observation. Simulation work has suggested a canonical value of 1.7 for the color correction factor \citep{Shimura1995ApJ...445..780S, Davis2005ApJ...621..372D}. However, in principle, this factor should vary with spectral state and accretion rate \citep[e.g.,][]{Merloni2000MNRAS.313..193M, Ren2022ApJ...932...66R}.

\subsubsection{Reflection Spectroscopy}

\begin{figure}[h]
    \centering
    \includegraphics[width=0.98\linewidth]{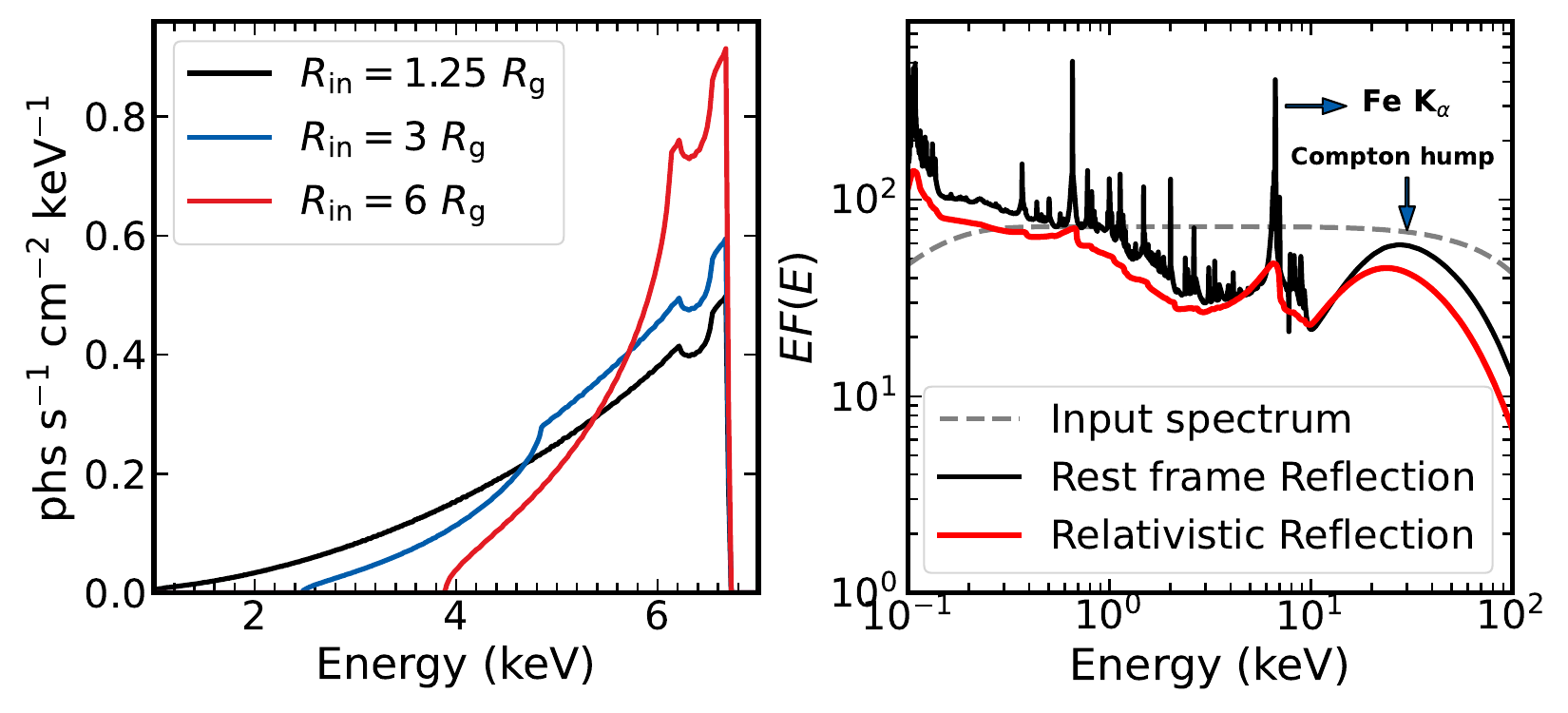}
    \caption{(Left) The broad iron line profiles as a function of the inner disk radius ($R_{\rm in}$). The profiles 
    are calculated using the \texttt{relline} model assuming a black spin $a_*=0.998$, 
    an inclination angle $i=30^{\circ}$, 
    an power-law emissivity profile with the index $q=3$ and an outer disk radius $R_{\rm out}=400~R_{\rm g}$.
    (Right) Rest-frame reflection (black) and relativistic reflection (red) spectra for an accretion
    disk extending to the innermost circular orbit. 
    }
    \label{line_ref}
\end{figure}

Analysis of the reflection component is also a powerful tool for studying the accretion geometry. 
As shown in Figure~\ref{line_ref}, the reflection spectrum in the local rest frame of the disk is characterized by 
fluorescent emission lines (the strongest 
one being the iron K${\alpha}$ line), photoelectric absorption edges and a Compton reflection hump 
peaking around 30~keV \citep{George1991,Garcia2010}. The commonly used local reflection models are 
\texttt{reflionx}\footnote{\url{https://github.com/honghui-liu/reflionx_tables}} \citep{Ross2005} and \texttt{xillver} \citep{Garcia2010}. These models calculate the 
reflection spectrum of an ionized disk. There are also models specifically for neutral disks, such as 
\texttt{pexrav}\footnote{\url{https://heasarc.gsfc.nasa.gov/xanadu/xspec/manual/node214.html}} 
\citep{Magdziarz1995MNRAS.273..837M} and 
\texttt{pexmon}\footnote{\url{https://heasarc.gsfc.nasa.gov/xanadu/xspec/manual/node213.html}} 
\citep{Nandra2007MNRAS.382..194N}.

The local reflection spectrum, resulting from irradiation of the disk by the corona, 
will be altered by several relativistic effects, including Doppler shifts, 
gravitational redshift, and light bending. The 
left panel of Figure~\ref{line_ref} illustrates how a narrow line at 6.4~keV becomes broadened and skewed 
due to these effects from the accretion disk when integrated
over the whole profile of the disk. A detailed discussion of how various system parameters (e.g., inner disk radius, disk inclination angle, etc.) impact 
the broad line profile can be found in \cite{Fabian2000} and \cite{Gates2024arXiv241114338G}. 
The red wing of the broadened profile extends to lower energy as the inner disk radius decreases, 
making it a useful probe of disk geometry. Additionally, the emissivity profile, namely the radial dependence of the 
reflected emission intensity, is determined by the coronal geometry 
\citep[e.g.,][]{Wilkins2011MNRAS.414.1269W,Wilkins2012MNRAS.424.1284W}, though it can be challenging to reconstruct the 
corona geometry directly from the observed phenomenological emissivity profile.
Several models exist to compute relativistic line broadening, including \texttt{diskline} \citep{Fabian1989}
\texttt{laor} \citep{Laor1991ApJ...376...90L}, \texttt{kyrline} \citep{Dovvciak2004ApJS..153..205D}, 
\texttt{kerrdisk} \citep{Brenneman2006ApJ...652.1028B} and \texttt{relline} \citep{Dauser2010}.

Applying the broadening kernel to the local reflection spectrum yields the relativistic reflection 
model \citep[e.g.,][]{Bambi2024Univ...10..451B}, 
which can be used to fit observational data and infer the disk-corona geometry. A summary of available 
reflection models and possible source of systematic uncertainties can be found in \cite{Bambi2021}. 
The two main features of the resulting relativistic
reflection spectrum seen by an external observer are the broad iron line 
around 6.4~keV and the Compton hump peaked around 30~keV. 
These features have been commonly found in the hard and intermediate states of BH XRBs 
\citep[e.g.,][]{Walton2016, Jiang2019gx339, Liu2023ApJ...950....5L, Liu2023ApJ...951..145L}.

X-ray reflection models have been significantly developed over the past 15 years. However, like all models, they rely on several simplifications. These can introduce systematic uncertainties into parameter estimation, in addition to statistical uncertainties. The key simplifications can be classified into the following categories: 

\begin{itemize}
    \item \textbf{\textit{Simplifications in disk geometry assumptions.}} The accretion disk is often modeled as an infinitesimally thin Keplerian disk aligned perpendicularly to the black hole spin axis. However, a realistic disk has a finite thickness \citep{Taylor2018ApJ...855..120T}. The impact of this thin-disk assumption on black hole spin measurements (thus $R_{\rm in}$) is marginal in the sub-Eddington regime (e.g., $L/L_{\rm Edd}<0.3$; \citealt{Abdikamalov2020ApJ...899...80A, Tripathi2021ApJ...913..129T, Jiang2022MNRAS.514.3246J}). In the near-Eddington regime, radiation pressure causes vertical expansion of the disk, resulting in a thick-disk geometry rather than a thin one. Simulations have shown that applying the thin-disk assumption under these conditions can lead to significant biases in the estimation of key parameters like BH spin \citep{Taylor2018ApJ...855..120T, Riaz2020ApJ...895...61R, Riaz2020MNRAS.491..417R}. In the super-Eddington regime, strong, optically thick outflows with relativistic velocities and large opening angles become the primary component responsible for reflecting the coronal emission \citep{Thomsen2019ApJ...884L..21T}. Consequently, line broadening is dominated by the kinematics of the wind and its opening angle \citep{Zhang2024ApJ...977..157Z}. Therefore, the thin-disk assumption must be significantly modified to fit observational data in this regime \citep[e.g.,][]{Shashank2024arXiv240712890S}.
    \item \textbf{\textit{Simplifications in coronal geometry assumptions.}} The coronal geometry determines the emissivity profile of the disk reflection component \citep{Wilkins2012MNRAS.424.1284W}, but it remains poorly constrained. Simple geometries, such as the lamppost model where the corona is approximated as a point source located at a certain height above the black hole, have been used to construct reflection models \citep{Dauser2013}. However, a realistic corona much have a radial extension to efficiently intercept seed photons from the disk. Therefore, models implementing a disk-like or ring-like corona above the BH have also been developed \citep{Riaz2022ApJ...925...51R}. 
    The ring-like coronal model has been applied to an AGN system and described the data well \citep{Nekrasov2025A&A...704A.129N}. They found that the corona should be compact, with both its vertical and radial extents being less than $\sim 3~R_{\rm g}$. This is in agreement with what is implied by fitting the data using a simple lamppost model. Such a test proves the concept of developing and applying more realistic coronal geometries for X-ray data.
    Alternatively, it is common to model the emissivity profile directly using phenomenological models, such as a power-law or broken power-law. In practice, it is recommended to test different geometries during spectral fitting to check for model dependence in the results.
    \item \textbf{\textit{Simplifications in the calculation of local reflection spectrum.}} Current reflection models are calculated assuming a parallel-slab configuration with a constant density in the vertical direction. This constant-density assumption might be appropriate in the radiation pressure-dominated regime. However, the accretion disk in the hard state of BH XRBs is expected to be gas pressure-dominated, a condition that naturally predicts a vertically stratified density structure, with higher density towards the disk mid-plane. Evidence for such structure has been presented by \cite{Liu2023ApJ...951..145L}, who found that the densities measured at the disk surface in the hard state of BH XRBs are lower than the theoretical predictions for the disk center.
    Local reflection spectra have been calculated for a gaussian density profile \citep{Ballantyne2001MNRAS.327...10B} or a density profile from hydrostatic equilibrium \citep{Nayakshin2000ApJ...537..833N}. However, a full model that can be tested on data is still lacking. Moreover, most currently employed reflection models often assume no upward radiation flux at the bottom boundary. This is appropriate for AGN disks, which are relatively cold. In contrast, disks in XRBs can be significantly hotter, with temperatures reaching 1--2 keV in the soft and intermediate states. This intrinsic thermal emission can therefore influence the ionization structure of the disk. The impact of a hot disk has been incorporated in the \texttt{reflionx} model \citep{Ross2007} and studied in specific sources \citep{Reis2008, Reis2009, Reis2011, Reis2012ApJ...751...34R, Steiner2011MNRAS.416..941S, Steiner2012MNRAS.427.2552S, Walton2012MNRAS.422.2510W, Chiang2012MNRAS.425.2436C, Reis2013ApJ...778..155R, King2014ApJ...784L...2K, Xu2020ApJ...893...30X}. A test of the model by \cite{Liu2023ApJ...950....5L} found no strong impact on the measurement of $R_{\rm in}$. A caveat even with these studies, however, is the assumption of a single-temperature blackbody spectrum for the disk. In reality, an accretion disk emits a multi-temperature blackbody spectrum, with higher temperatures at smaller radii. This radial temperature gradient will inevitably lead to a radius-dependent ionization state.
\end{itemize}


\subsection{X-ray timing}
\label{timing}

The accretion flow is highly dynamic, exhibiting X-ray variability on timescales ranging from milliseconds to years. 
Timing analysis mainly operates in the frequency domain. One of the main timing products is the power spectral density 
(PSD), which shows the variability power as a function of frequency \citep{vanderKlis1989ASIC..262...27V, Uttley2014A&ARv..22...72U}. 
In the hard and intermediate states, the PSD often shows narrow peak features 
known as quasi-periodic oscillations (QPOs), indicating enhanced variability at characteristic frequencies 
\citep{Motta2011MNRAS.418.2292M, Ingram2019NewAR..8501524I, Liu2021ApJ...909...63L}. 
The QPO frequencies may be linked to dynamic processes in the system (e.g., 
Lense-Thirring precession, \citealt{Ingram2009MNRAS.397L.101I}) or to oscillation modes of the accretion flow 
\citep{Tagger1999A&A...349.1003T, Varniere2002A&A...394..329V, Chakrabarti2008A&A...489L..41C, Varniere2012A&A...545A..40V}. Therefore, 
they may potentially provide insights into the accretion geometry or fundamental parameters 
such as the black hole mass and spin \citep[e.g.,][]{Motta2014MNRAS.437.2554M}. 

\begin{figure}[h]
    \centering
    \includegraphics[width=0.49\linewidth]{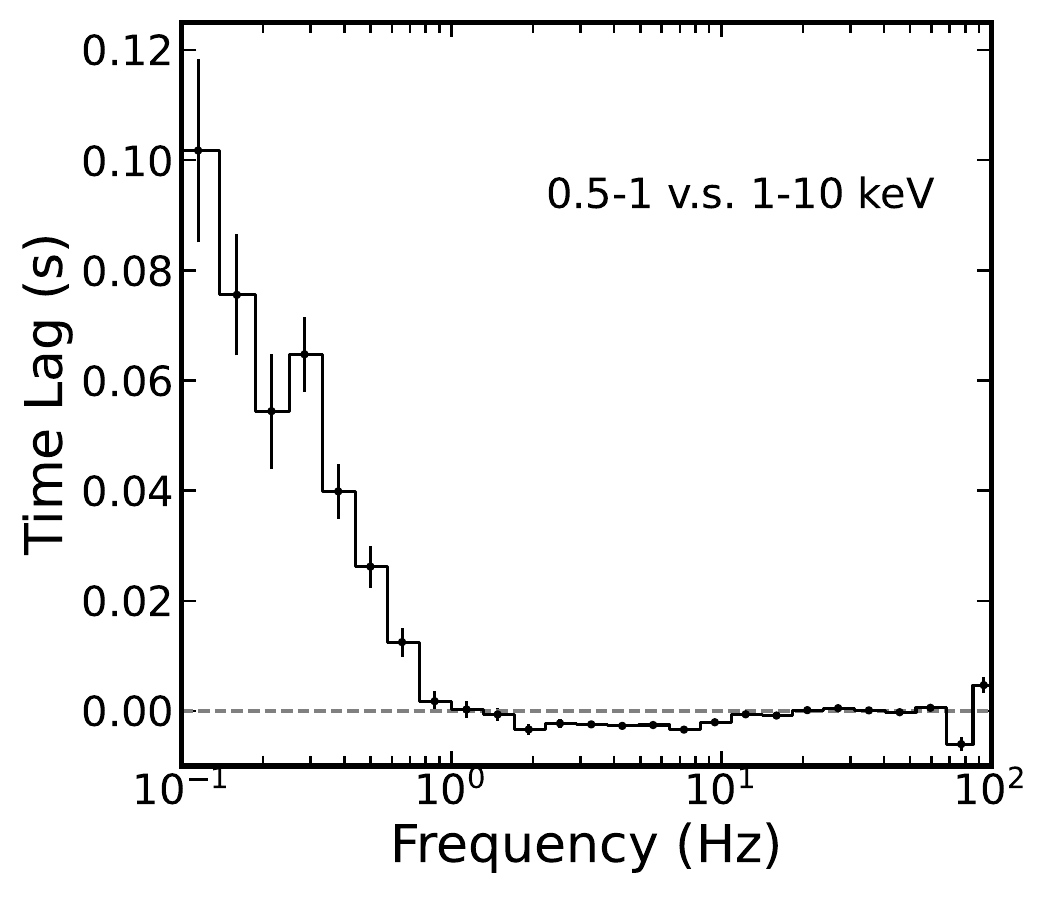} 
    \includegraphics[width=0.49\linewidth]{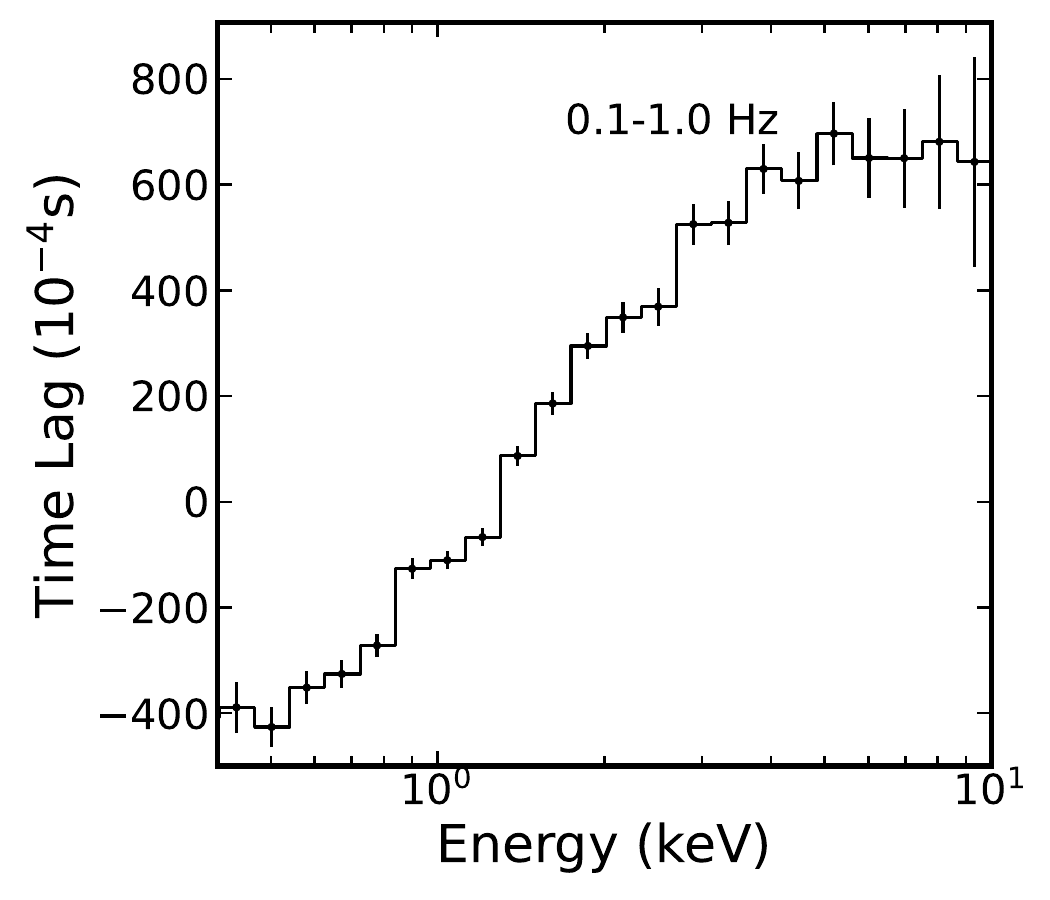} \\
    \includegraphics[width=0.49\linewidth]{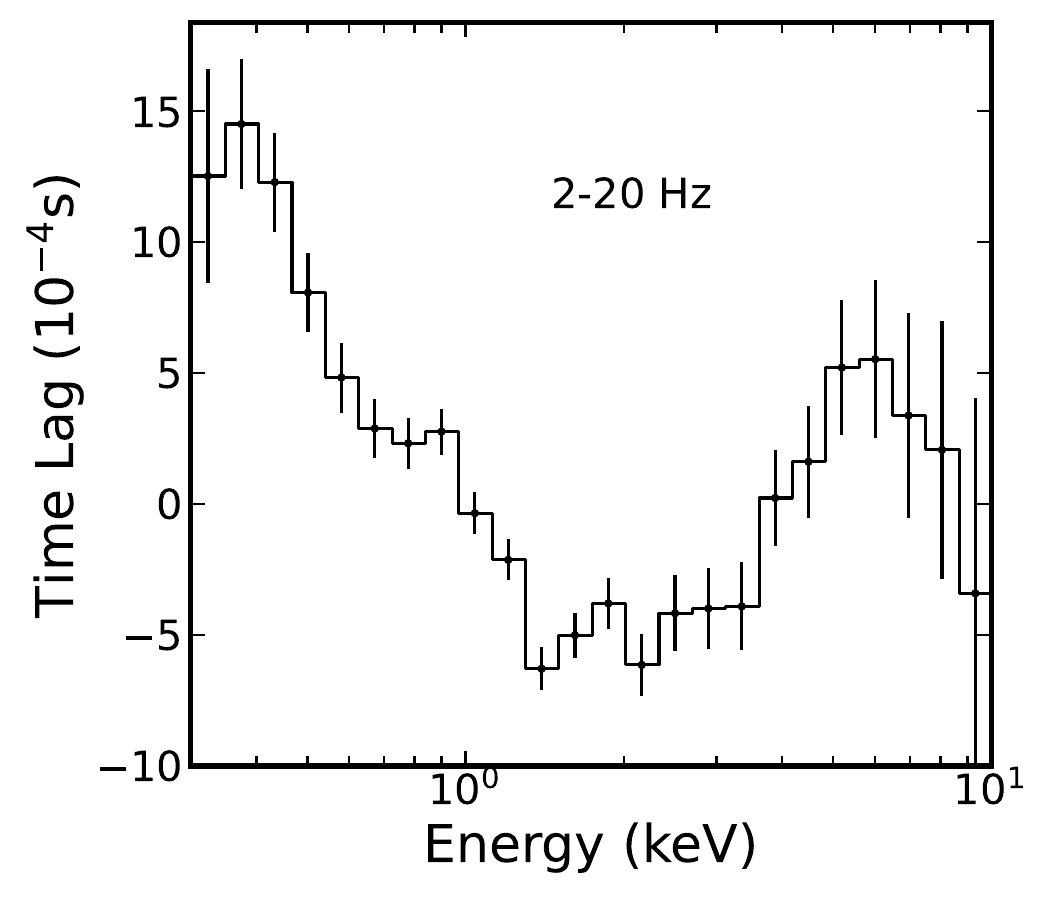}
    \caption{(Upper Left) Lag-frequency spectrum between the 0.5--1 keV and 1--10 keV bands for the black hole X-ray binary MAXI J1820+070 from a NICER observation. (Upper Right) Lag-energy spectrum from the same data in the 0.1-1.0 Hz frequency range, using the 0.5--10 keV band as a reference. (Lower) Lag-energy spectrum in the 2-20 Hz frequency range.
    }
    \label{Lag_pro}
\end{figure}

Other important timing products are the lag-frequency spectrum (e.g., 
the phase lag between lightcurves of two energy bands) and the lag-energy spectrum (phase lag 
relative to a reference energy band in a specific frequency range). 
A typical lag-frequency spectrum (0.5--1.0 v.s. 1.0-10.0 keV) for a BH XRB is shown in the left panel of Figure~\ref{Lag_pro}. 
There is strong contribution from the reflection component in the soft band and the 
hard band is dominated by the coronal emission.
A hard/positive lag (i.e., hard photons lagging behind soft photons) is observed at 
low frequencies (0.1--1.0 Hz), and is thought to originate 
from mass accretion rate fluctuations propagating through the accretion disk 
\citep{Lyubarskii1997MNRAS.292..679L, Ingram2011MNRAS.415.2323I, Ingram2012MNRAS.419.2369I}. 
In the lag-energy spectrum at low frequencies, the lag continues to rise with energy (the \textit{continuum lag}, see the right panel of Figure~\ref{Lag_pro}). In this spectrum, one should focus on the relative shape because the absolute lag values depend on the choice of the reference band. The lag-energy spectrum should be interpreted from bottom to top; that is, a smaller lag indicates that the signal arrives earlier at the detector. The steadily increasing trend with energy implies that high-energy
X-ray photons always lag behind low-energy photons when low-frequency variations are considered.
At high frequencies, the \textit{reverberation lag}, which has a more complex structure, starts to dominate. This lag arises because the reflection component responds to variability in the coronal emission with a delay due to light-travel time \citep[e.g.][]{Zhan2025arXiv250203995Z}. Consequently, the lag-energy spectrum at high frequencies resembles the shape of the reflected energy spectrum (see the lower panel of Figure~\ref{Lag_pro}).


It is evident that the tool can be applied to both XRB and AGN systems. However, most reverberation mapping studies over the past two decades have focused on AGNs \citep{Fabian2009Natur.459..540F, Zoghbi2012MNRAS.422..129Z, Zoghbi2014ApJ...789...56Z, Kara2015MNRAS.446..737K, Kara2016MNRAS.462..511K, Mastroserio2020MNRAS.498.4971M}, as detecting reverberation in XRBs is more challenging due to their smaller lag amplitudes and higher frequency ranges. A comparison of lag properties between XRBs and AGNs can be seen in Figure 7 of \cite{Kara2025ARA&A..63..379K}, which illustrates how the lag amplitude scales with BH mass and how the reverberation-dominated frequencies scale inversely with it.

One commonly used model for analyzing reverberation lags 
is \texttt{reltrans}\footnote{\url{https://github.com/reltrans}} 
\citep{Ingram2019MNRAS.488..324I}, which can simultaneously fit the reflection spectrum. The model assumes a 
lamppost geometry for the corona and can constrain its height using both spectral and timing information. Another important 
model in the literature is \texttt{vKompth}\footnote{\url{https://github.com/candebellavita/vkompth}} 
\citep{Bellavita2022MNRAS.515.2099B}, which is primarily designed for the
study of QPOs and fits the energy-dependent rms-amplitude and 
phase-lag spectra at the QPO frequency to estimate the size of a spherical corona.
One strength of the \texttt{vKompth} model is that it can fit the time-averaged Comptonization spectrum simultaneously with the lag and rms spectra. This model attributes the hard lag to the Comptonization process, where higher-energy photons require more scatterings within the corona. Note that this is a different assumption from the fluctuation-propagation scenario. The soft lag is attributed to the feedback and reprocessing of coronal photons in the disk, which is treated as thermalization into a blackbody spectrum. The relative importance of these two processes is controlled by the feedback parameter, defined as the fraction of the disk thermal emission that originates from reprocessed coronal emission. This parameter can be converted into the ``intrinsic feedback", which is the fraction of the coronal emission that returns to the disk.

\subsection{X-ray Polarimetry}

The measurement of X-ray polarization properties, i.e., the polarization degree (PD) and polarization angle (PA), 
offers strong potential for probing the details of the disk-corona geometry. These two parameters represent how electric 
field vectors of X-ray photons are preferentially distributed along a specific orientation. 
They are sensitive to the geometry of the emission region and the underlying emission mechanism. 

In the 
case of the thermally Comptonized emission, a higher PD is expected in more asymmetric systems, while the PA 
indicates the orientation of the Comptonizing medium \citep{Schnittman2010ApJ...712..908S}. 
The PD, PA, and their energy dependence for a given coronal geometry can be calculated using Monte Carlo-based codes such as \texttt{MONK} \citep{Zhang2019ApJ...875..148Z} and \texttt{KerrC} \citep{Krawczynski2022ApJ...934....4K}. These codes can also provide the simultaneous energy spectrum of the corona. However, it can be computationally expensive to construct a grid of models to fit spectro-polarimetric data. Alternatively, for a slab geometry, these polarimetric properties can be calculated using a radiation transport solver such as \texttt{COMPPS} \citep{Poutanen1996ApJ...470..249P, veledina_2022_7116125}.

The thermal emission from an optically thick accretion disk is expected to be polarized parallel 
to the disk plane, with a PD that depends on the inclination angle 
\citep{Chandrasekhar1960ratr.book.....C}. When considering scattering as the only relevant process, the polarization properties of a disk with finite optical depth can be calculated using the Monte Carlo code \texttt{STOKES}\footnote{The \texttt{STOKES} code is available at: \url{https://people.astro.unistra.fr/marinf/STOKES_web/index.html}. Pre-calculated polarimetric tables for disk thermal emission can be downloaded from: \url{https://projects.asu.cas.cz/stronggravity/kyn\#kynbb}.} \citep{Goosmann2007A&A...465..129G, Marin2018A&A...615A.171M}. Since the disk geometry can be approximated as a slab, its polarimetric properties for various optical depths can be pre-calculated and tabulated efficiently. A more self-consistent treatment should consider other processes such as absorption. This can be achieved by combining the \texttt{STOKES} code with a radiative transfer and photoionization code that can calculate the ionization structure of the disk surface \citep[e.g.,][]{Taverna2021MNRAS.501.3393T}. General relativistic (GR) effects are also crucial, as polarization vectors rotate due to parallel transport along null geodesics. This effect is more significant for larger BH spins, offering an alternative method for constraining the spin parameter \citep{Dovvciak2008MNRAS.391...32D}. Spectro-polarimetric models for disk thermal emission that incorporate these effects have been developed, such as \texttt{KYNBB} \citep{Dovvciak2008MNRAS.391...32D} and \texttt{KYNBBRR} \citep{Taverna2020MNRAS.493.4960T}. The latter model additionally accounts for the returning radiation effect, whereby photons emitted from the disk are bent by the strong gravitational field and fall back onto the disk.

The reflection component also 
has distinct polarization signatures, including a highly polarized continuum and depolarized 
fluorescent lines \citep{Matt1993MNRAS.260..663M, Podgorny2022MNRAS.510.4723P, Podgorny2023MNRAS.524.3853P}.
The exact polarization spectrum depends on the coronal geometry and its polarimetric properties, the ionization state of the disk, the inclination angle, and the black hole spin. A spectro-polarimetric model, \texttt{KYNSTOKES} \citep{Podgorny2023MNRAS.524.3853P}, has been developed to account for these effects. In this model, the spectral and polarimetric properties of the local disk reflection are calculated by combining the radiative transfer code \texttt{TITAN} \citep{Dumont2003A&A...407...13D} with the \texttt{STOKES} code. General relativistic effects are subsequently incorporated using the \texttt{KY} package \citep{Dovvciak2004ApJS..153..205D, Dovvciak2011ApJ...731...75D}.

High sensitivity X-ray polarimetry measurements have only been possible since the launch of the 
Imaging X-ray Polarimetry Explorer (IXPE) on 9 December 2021 \cite{Weisskopf2022JATIS...8b6002W}. 
IXPE, operating in the 2--8 keV band, has observed a dozen BH XRBs, yielding significant breakthroughs 
in the understanding of the disk-corona geometry \citep[see][for a recent review]{Dovvciak2024Galax..12...54D}.

\section{Accretion geometry inferred from X-ray observations}
\label{acc_geo}

As our main vehicle for summarising our current
understanding of the evolving accretion geometry in BH XRBs, a compilation
of inner disk radius ($R_{\rm in}$) measurements from pile-up\footnote{Pile-up is an effect where two 
or more photons hit the same pixel (or neighboring pixels) during a single readout frame. The detector will will incorrectly register them as one single event, distorting the observed X-ray spectrum.} free data is shown in 
Figure~\ref{Rin}. We select sources that meet the following three criteria:
(1) Available distance and BH mass measurements to compute the Eddington limit.
(2) Eddington ratio measurements (e.g., $L_{\rm X}/L_{\rm Edd}$) published alongside the $R_{\rm in}$ measurements using relativistic models.
(3) Data obtained from pile-up free instruments, such as \textit{RXTE}, \textit{Insight}-HXMT, \textit{NuSTAR}, 
and \textit{NICER}, in the hard and intermediate states. Basic information about the sources can be found in Table~\ref{sources}.
In cases where the same observation has been analyzed multiple times in the literature, we include only the most recent measurement of $R_{\rm in}$. 
These measurements are derived from X-ray spectroscopy, as current X-ray timing and 
polarimetry data are not yet capable of reliably constraining the $R_{\rm in}$ parameter for XRBs. 
The $R_{\rm in}$ measurements in Figure~\ref{Rin} are predominantly obtained via the reflection method. A notable exception is MAXI~J1820+070, where the values are derived from a combination of continuum fitting and reflection \citep{Fan2024ApJ...969...61F} (excluding the two lowest-luminosity points from \citealt{Xu2020ApJ...893...42X}). 
We will discuss the 
accretion geometry based on Figure~\ref{Rin}, incorporating relevant insights from X-ray timing and polarimetry 
measurements. 

\subsection{Hard state (Region A to B)}

\begin{figure}[h]
    \centering
    \includegraphics[width=0.98\linewidth]{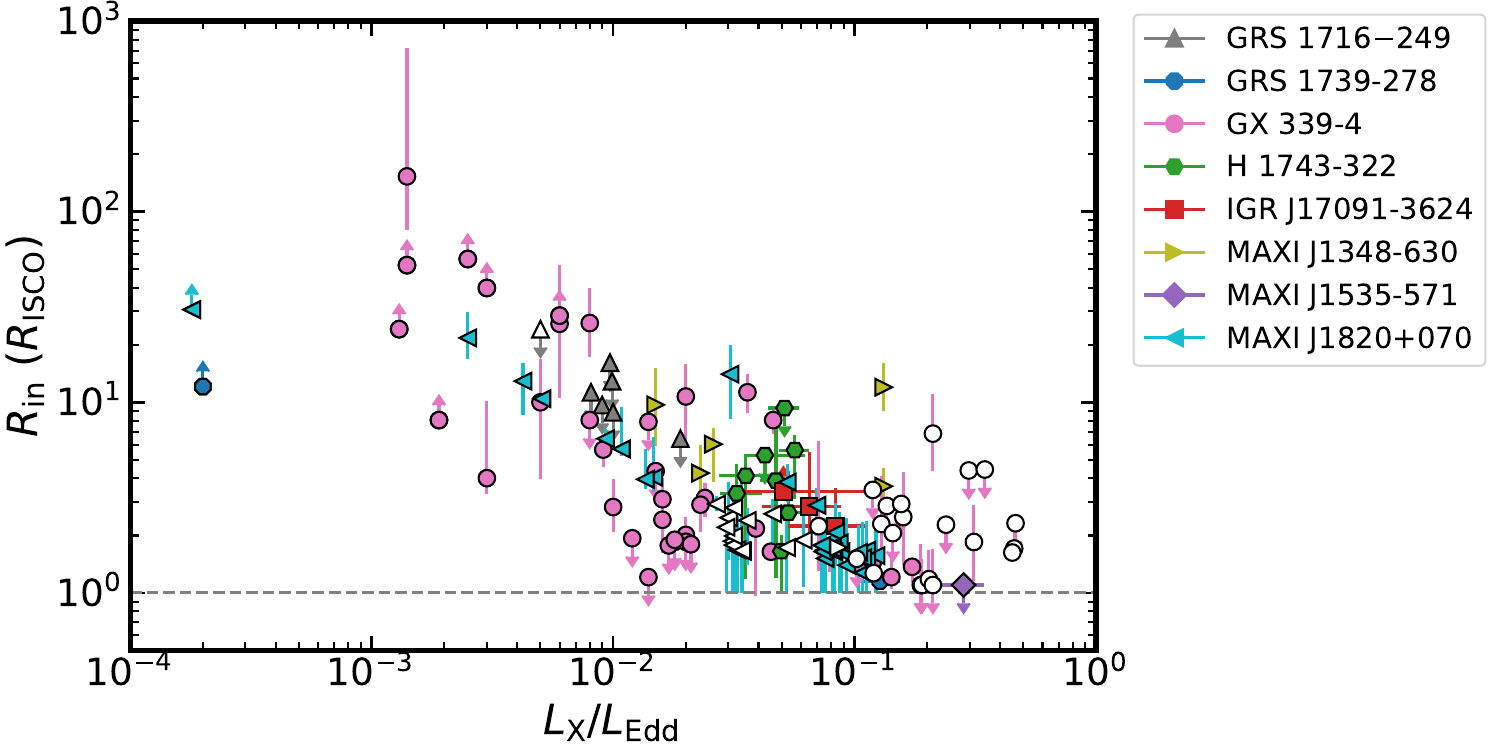}
    \caption{The evolution of the inner disk radius ($R_{\rm in}$) with the Eddington-scaled luminosity. 
    The measurements of $R_{\rm in}$ are from X-ray spectroscopy. The filled symbols represent 
    hard state data and the empty symbols are for intermediate states. Information of the selected sources can be found in Table~\ref{sources}.
    \newline
    \textbf{(References)} GX~339--4: \cite{Reis2008}, \cite{Tomsick2008ApJ...680..593T}, \cite{Tomsick2009},
    \cite{Shidatsu2011}, \cite{Petrucci2014}, \cite{Garcia2015}, \cite{Wang2018},
    \cite{Garcia2019ApJ...885...48G}, \cite{Sridhar2020}, \cite{Wang2020ApJ...899...44W}, \cite{Liu2023ApJ...950....5L};
    GRS~1739--278: \cite{Furst2016ApJ...832..115F}, \cite{Liu2023ApJ...951..145L}; 
    GRS~1716--249: \cite{Jiang2020MNRAS.492.1947J}; 
    MAXI~J1348--630: \cite{Jia2022MNRAS.511.3125J};
    IGR~J17091--3624, H~1743--322, MAXI~J1535--571: \cite{Liu2023ApJ...951..145L} and MAXI~J1820+070: \cite{Xu2020ApJ...893...42X}, \cite{Fan2024ApJ...969...61F}.
    }
    \label{Rin}
\end{figure}

This region corresponds to the right branch of the HID and represents the emergence from quiescence 
and the evolution from the low hard (region A of Figure~\ref{hid_spec}) to the bright hard (B) states. 
Figure~\ref{Rin} shows that at the beginning and end of the outburst 
(region A), when the accretion rate is 
low ($L_{\rm X}/L_{\rm Edd}\sim0.1\%$)\footnote{$L_{\rm Edd}$ is the Eddington luminosity defined as 
$L_{\rm Edd}=1.26\times 10^{38}~M_{\rm BH}/M_{\odot}$~erg~s$^{-1}$.},
$R_{\rm in}$ is truncated at large radii (e.g., $>100~R_{\rm ISCO}$). As the accretion rate increases, 
the disk moves closer to the BH, but typically remains truncated at $\sim 10~R_{\rm g}$ 
if $L_{\rm X}/L_{\rm Edd}<$ 1-2\%. In this regime, a hot inner flow likely exists between the inner edge of the disk 
and the BH, acting as the corona. This radially extended coronal geometry in the low-hard state is consistent with IXPE measurements of Cygnus~X-1 \citep[taken at $L_{\rm X}/L_{\rm Edd}\sim 1\%$,][]{Krawczynski2022Sci...378..650K} and 
Swift~J1727.8--1613 \citep[taken at $L_{\rm X}/L_{\rm Edd}\sim 0.5\%$,][]{Podgorny2024A&A...686L..12P}. 
In both cases, the PA of the coronal emission was found to align with the radio jet direction, suggesting 
a radially extended corona (i.e. extended in the disk plane).
IXPE measurements also rule out the lamppost and spherical lamppost coronal geometries, as they produce a PD that is too low compared to the observed values \citep{Ursini2022MNRAS.510.3674U, Krawczynski2022Sci...378..650K}.
The decreasing trend of $R_{\rm in}$ with luminosity along the right branch is also consistent with the X-ray reverberation measurements. 
For instance, \cite{Wang2022ApJ...930...18W} found that the reverberation lag amplitude in eight BH XRBs decreases along the right 
branch (see their Figure 6), which would naturally result from the disk moving inward toward the BH and the corona 
contracting radially. 

As the accretion rate continues to increase, $R_{\rm in}$ can reach the ISCO even in the hard state 
when $L_{\rm X}/L_{\rm Edd}>$ 3-4\%. Among all the hard-state measurements in Figure~\ref{Rin} with 
$L_{\rm X}/L_{\rm Edd} > 3\%$ (23 in total), 78\% (18/23) are consistent with $R_{\rm in} < 4~R_{\rm ISCO}$, and 
43\% (10/23) are consistent with $R_{\rm in} < 2~R_{\rm ISCO}$ (including three upper limits). These 
findings indicate that 
\textit{the accretion disk is not necessarily truncated in the bright hard state}.
The geometric evolution of the disk-corona system already takes place in the hard state and is not limited to the state transition. This is not surprising given that the source luminosity varies by orders of magnitude during this phase.

An interesting finding is that even when the disk
reaches (and remains at) the ISCO during the bright hard state (region B), 
the reverberation lag amplitude continues to decrease and the upper bound of the soft lag frequency\footnote{This is 
the frequency ($f_0$) at which the lag-frequency spectrum crosses zero. The lag flips into positive at this frequency due to phase wraping. Therefore, this frequecy is related to the intrinsic time delay ($\tau$) between the coronal and reflected photons in a way: $f_0 \sim 1/2\tau$} tends 
to increase with time \citep[see][]{Kara2019Natur.565..198K,Wang2022ApJ...930...18W}. Since the inner radius of the disk is stable in this region, 
the decreasing lag and increasing frequency are interpreted as the corona shrinking in the vertical direction. This scenario 
is further supported by spectral fitting of the \textit{NuSTAR} data of MAXI~J1820+070 using a dual-lamppost 
model \citep{Buisson2019}, which can mimic a vertically extended corona \citep{Lucchini2023ApJ...951...19L}.

There are hints that indicate a connection between the corona and the jet.
It has been suggested that the base of the jet could produce the Comptonized emission \citep{Markoff2005ApJ...635.1203M, Cao2022MNRAS.509.2517C}.
In the bright hard state of MAXI~J1820+070, the reflection strength has been found to decrease with time while the iron line profile remains stable. The stable line profile indicates a constant inner disk radius. This decreasing reflection strength has been interpreted with a scenario involving a jet-like corona whose bulk velocity increases with time \citep{You2021NatCo..12.1025Y}. In the same MAXI~J1820+070 data, \cite{Ma2021NatAs...5...94M} detected low-frequency QPOs above 200~keV. They also discovered a long soft lag ($\sim$0.9~s) at the QPO frequency between 150--200 keV photons and those below 30~keV. This was interpreted within a precessing jet scenario, where high-energy photons originate from the jet base and low-energy photons from farther out in the jet. The jet's precession produces the QPO, and a soft lag would naturally arise if the jet base is observed before the downstream emission region.
Notably, the outflowing corona scenario has also been invoked to explain the higher-than-expected PDs 
observed in \textit{IXPE} observations of several BH XRBs \citep{Krawczynski2022Sci...378..650K, Poutanen2023ApJ...949L..10P, 
Ratheesh2024ApJ...964...77R, Ewing2025MNRAS.541.1774E}. In the hard state of Cygnus~X-1, \textit{IXPE} measured a PD $\sim$ 4\%, which is higher than the expected PD of 1\% given the orbital inclination of the system \citep{Krawczynski2022ApJ...934....4K}. Similarly, a high PD of 9\% was found in the hard state of IGR J17091--3624, which is high even for a high-inclination system \citep{Ewing2025MNRAS.541.1774E}. An outflowing corona with a non-zero bulk velocity could increase the PD due to relativistic aberration \citep{Poutanen2023ApJ...949L..10P}, offering a potential explanation for these high polarization measurements. For Cygnus~X-1, an alternative explanation for the high observed PD is that the corona is more inclined than the binary system, i.e., a warped disk. Such a configuration is possible if the orbital angular momentum is not aligned with the BH spin \citep{Bardeen1975ApJ...195L..65B}.

\begin{table*}
    \centering
    \caption{Selected sources with distance, mass, and inclination measurements.}
    \label{sources}
    \renewcommand\arraystretch{1.5}
    \begin{tabular}{lcccc}
        \hline\hline
        Source           & Distance (kpc)       & Mass ($M_{\odot}$) & Inclination ($^{\circ}$) & Ref \\
        \hline
        GRS~1716--249   & $6.9\pm1.1$   & $6.4_{-2.0}^{+3.2}$  & $61\pm15$ (binary)  & 1  \\
        \hline
        GRS~1739--278    & 6--8.5 & 4.0-9.5 & -  & 2,3 \\
        \hline
        GX~339--4  & 8--12  & 4--11 & 40--60 (binary)  & 4 \\
        \hline
        H~1743--322      & $8.5\pm0.8$ & $12\pm2$ & $75\pm3$ (jet) & 5,6 \\
        \hline
        IGR~J17091--3624 & 11--17 & 8.7--15.6 & - & 7,8 \\  
        \hline
        MAXI~J1348--630  & $3.39\pm0.38$ & $11\pm2$  &  - & 9 \\
        \hline
        MAXI~J1820+070   & $2.96\pm0.33$ & $9.2\pm1.3$ & $63\pm3$ (jet) & 10  \\
        \hline
        MAXI~J1535--571  & $4.1_{-0.5}^{+0.6}$  & $10.4\pm0.6$ & - & 11,12  \\
        \hline
    \end{tabular} \\

    \textit{Note.} Selected sources and their properties. For references about the measurements of the distance, mass and inclination angle: (1) \cite{Casares2023MNRAS.526.5209C}; (2) \cite{Greiner1996}; (3) \cite{Wang2018PASJ...70...67W}; (4) \cite{Zdziarski2019}; (5) \cite{Steiner2012}; (6) \cite{Nathan2024MNRAS.533.2441N}; (7) \cite{Rodriguez2011}; (8) \cite{Iyer2015}; (9) \cite{Lamer2021A&A...647A...7L}; (10) \cite{Atri2020MNRAS.493L..81A}; (11) \cite{Chauhan2019}; (12) \cite{Sridhar2019MNRAS.487.4221S} 
\end{table*}

\subsection{Intermediate states (Region B to C)}

This region corresponds to the upper branch of the HID and marks the hard-to-soft state transition. 
Since the accretion disk already extends close to the ISCO radius before the transition 
(i.e., during the bright hard state), we expect $R_{\rm in}$ to remain near $R_{\rm ISCO}$ 
throughout the state transition. Monitoring the transition process is challenging, as the starting point is unpredictable, and 
the transition timescale is short. The best-monitored source is GX~339--4, for which hard-to-soft transitions 
have been captured multiple times by \textit{RXTE} \citep{Sridhar2020} and \textit{Insight}--HXMT \citep{Liu2023ApJ...950....5L}.
The spectroscopic data from GX~339--4 indeed suggest that $R_{\rm in}$ remains stable and close to $R_{\rm ISCO}$ during 
the state transition (see empty symbols in Figure~\ref{Rin}). Spectroscopic studies of other sources have also confirmed this behavior
\citep[e.g.,][]{Tao2019ApJ...887..184T, Draghis2020ApJ...900...78D, Wang2021, 
Ren2022ApJ...932...66R, Fan2024ApJ...969...61F, Adegoke2024ApJ...977...26A}. Notably, the disk temperature at $R_{\rm in}$ continues to increase toward the soft state \citep{Liu2023ApJ...950....5L}, which could provide stronger cooling to the corona.

Spectroscopic studies can suffer from degeneracies. Such degeneracies prevent us from 
placing strong constraints on the geometry and 
dynamics of the corona. However, combining spectroscopy with timing techniques can yield additional insights. 
From the soft lag sample compiled by \cite{Wang2022ApJ...930...18W}, it was found that --- unlike the decreasing trend 
in the hard state --- the soft lag amplitude tends to increase during the hard-to-soft transition.
Fitting the lag spectra with the \texttt{reltrans} model further suggests that the corona height increases along with the transition \citep{Wang2021, DeMarco2021A&A...654A..14D}, indicating a vertical expansion of the corona that 
may be related to the transient jet ejections \citep[e.g.,][]{Davidson2025ApJ...994...54D}.
Similar corona-jet connections 
have also been inferred from correlations between radio flux and disk-corona parameters \citep[e.g.,][]{Vadawale2003ApJ...597.1023V, Mendez2022NatAs...6..577M}. Moreover, X-ray flaring associated with transient jet activity has been found in a few BH XRBs \citep{Punsly2013ApJ...764..173P, Walton2017}. In V404~Cyg, the jet itself appears to have temporarily acted as the X-ray source that illuminates the disk \citep{Walton2017}.

Note that an alternative explanation for the soft lag in the intermediate states has been proposed by \cite{Kawamura2023MNRAS.525.1280K} within the framework of mass accretion rate fluctuations. Typically, fluctuation propagation produces hard lags, as fluctuations from larger radii (emitting softer radiation) modulate the harder emission originating from smaller radii in a radially stratified corona. However, soft lags can also be produced if the hard Comptonization spectrum extends to lower energies than the soft Comptonization spectrum. This condition can be achieved if the seed photons for the central Comptonization region are much cooler than those from the inner disk edge that provide seed photons for the outer Comptonization region. Another potentially important effect is the thermalization/scattering time delay within the reflection process \citep{Salvesen2022ApJ...940L..22S}. This effect is more significant for softer emerging photons, as they must undergo numerous scatterings to thermalize before escaping the disk. Current reverberation models assume instantaneous reprocessing; however, the scattering time delay may be longer than the light-travel time delay in the intermediate states \citep{Salvesen2022ApJ...940L..22S}.

While the \texttt{reltrans} model does not apply to the lags at the QPO frequency, the \texttt{vKompth} 
model is specifically designed to interpret lag and RMS spectra associated with QPOs. 
This model has been applied to several sources, and the feedback parameter has been observed to decrease during the state transition compared to the hard state \citep[e.g.,][]{Garcia2021MNRAS.501.3173G, Garcia2022MNRAS.513.4196G, Zhang2022MNRAS.512.2686Z, Zhang2023MNRAS.520.5144Z, Ma2023MNRAS.525..854M, Peirano2023MNRAS.519.1336P}. Since this parameter represents the coupling strength between the disk and the corona, its decrease suggests that the corona evolves from a strongly coupled configuration (e.g., horizontally extended) to a weakly coupled one (e.g., vertically extended). However, it is important to note that the model inherently assumes a spherical coronal geometry. Therefore, this should be regarded as an empirical interpretation of the geometry, based on the data rather than a prediction of the model. In addition, the coronal sizes inferred from this model are generally large (hundreds of $R_{\rm g}$). In some cases, a two-corona scenario has been proposed to fully explain the QPO lag and rms spectra \citep{Peirano2023MNRAS.519.1336P}, in which the smaller corona is found to have a size of tens of $R_{\rm g}$ \citep[e.g.,][]{Ma2023MNRAS.525..854M, Alabarta2025ApJ...980..251A}. A spherical corona with a size of hundreds of $R_{\rm g}$
would be difficult to reconcile with constraints obtained from other methods. X-ray reverberation analyses of AGNs suggest that the corona is compact ($\sim$ 10 $R_{\rm g}$, \citealt{Emmanoulopoulos2014MNRAS.439.3931E}). Measurements from microlensing studies also indicate that the X-ray-emitting region in luminous AGNs has a size smaller than 10 $R_{\rm g}$
\citep{Dai2010ApJ...709..278D}. We expect a similar coronal size in BH XRBs, as these systems are essentially mass-scaled-down versions of AGNs \citep{McHardy2006Natur.444..730M, Reis2013ApJ...769L...7R}. Future model developments that consider alternative geometries, such as vertically extended or jet-like coronae, and more physical processes, such as the propagation of heating within the corona, may help to deliver a more consistent picture.

\begin{figure}[h]
    \centering
    \includegraphics[width=0.98\linewidth]{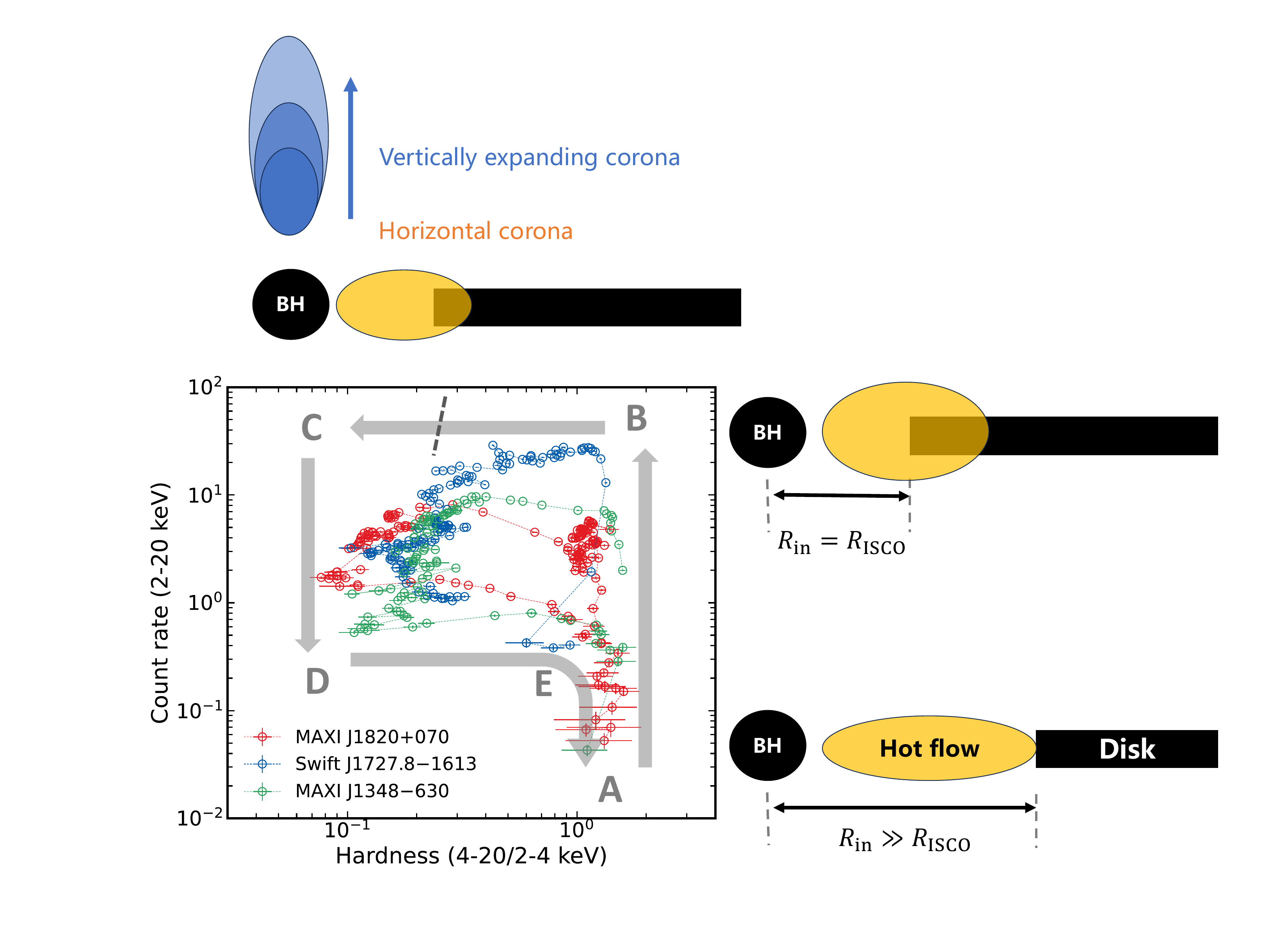}
    \caption{Sketch of hypothetic accretion geometries of BH XRBs along the hardness-intensity diagram.
    }
    \label{geometry}
\end{figure}

X-ray polarization measurements in this branch have been conducted during the hard-intermediate state, e.g., before 
the launch of discrete jets, of Swift J1727.8--1613 \citep{Veledina2023ApJ...958L..16V, Ingram2024ApJ...968...76I} 
and in the soft-intermediate of GX~339--4 \citep{Mastroserio2025ApJ...978L..19M}. In these observations, the PA 
is found to align with the jet direction, similar to the low-hard state. This may suggest that the corona is horizontally 
extended along the accretion disk plane, which appears to conflict with the vertically extended geometry implied 
by reverberation lag measurements and the small inner disk radius inferred from spectroscopic studies.
The latter conflict might be alleviated if the disk and corona are not radially separated, e.g., 
if the corona overlaps the disk (see Figure~\ref{geometry} and discussions in 
\citealt{Liu2022iSci...25j3544L}). The dual-corona structure proposed by 
\cite{Peirano2023MNRAS.519.1336P}, in which a horizontally extended corona dominates the spectral shape while a vertically 
extended corona dominates the lag, could potentially reconcile the former conflict. An illustration of such a dual-corona 
configuration is shown in the upper region of Figure~\ref{geometry}.
However, it should be noted that alternative scenarios also exist to explain 
the X-ray polarization measurements. For example, scattering off 
the walls of a jet could produce a PA aligned with the jet without requiring a horizontally extended corona 
\citep{Dexter2024MNRAS.528L.157D}. Alternatively, optically thin synchrotron emission from the jet could 
generate a similar PA, although its overall impact depends on its flux contribution to the IXPE band, 
the viewing angle, and the magnetic field structure \citep{Russell2014MNRAS.438.2083R}. This emission can 
be highly polarized (up to 70\%) and has already been detected in Swift J1727.8--1613 at 
gamma ray energies \citep{Bouchet2024A&A...688L...5B}. Lastly, such a PA could also arise from the returning 
radiation effect (i.e., photons emitted from the disk that return to the disk) due to the strong 
gravitational field near BHs \citep[e.g.,][]{Steiner2024ApJ...969L..30S}, or from scattering 
by disk winds \citep[e.g.,][]{Nitindala2025A&A...694A.230N}. Disentangling these scenarios 
requires more detailed multi-wavelength spectral, timing, and polarimetry measurements, as well as 
the development of simulations and data-fitting models.
If some of these alternative scenarios are at work, such as the scattering off the jet walls or disk winds, they may also contribute to the polarization signal in the hard state.

\subsection{Soft state (Region C to D)}

The soft state is believed to be described by the standard optically thick, geometrically
thin disk \citep{Shakura1973,Novikov1973blho.conf..343N}, with $R_{\rm in} = R_{\rm ISCO}$ \citep{Steiner2010} and 
low X-ray variability. Therefore, there are no X-ray reverberation results for the soft state. Compelling evidence supporting the standard disk scenario with $R_{\rm in} = R_{\rm ISCO}$, is the observed $L_{\rm disk} \propto T^{4}$ relation. This relation has been seen in the soft state for a number of sources where $L_{\rm disk}$ changes by nearly two orders of magnitude \citep{Kubota2001ApJ...560L.147K, Kubota2004ApJ...601..428K, Gierlinski2004}. This scaling indicates a constant inner disk radius, which is naturally identified with the ISCO.

Fitting relativistic disk thermal emission models to soft state X-ray spectra is one of the leading techniques to 
measure BH spin parameters \citep[continuum fitting,][]{Zhang1997ApJ...482L.155Z, Gou2011ApJ...742...85G, McClintock2014SSRv..183..295M}. 
This method typically assumes no emission within the ISCO radius (the plugging region). 
However, three-dimensional general relativistic magneto-hydrodynamic (GRMHD) simulations have shown that the plugging region 
could contribute with non-negligible emission to the total X-ray spectrum 
\citep[e.g.,][]{Zhu2012MNRAS.424.2504Z, Mummery2024MNRAS.532.3395M}. In several cases, emission from 
this region is found to be essential for adequately modeling soft-state X-ray spectra 
\citep[e.g.,][]{Fabian2020MNRAS.493.5389F, Mummery2024MNRAS.533L..83M}. This ``extra'' 
component may potentially affect continuum-fitting black hole spin measurements \citep[e.g.,][]{Mummery2024MNRAS.531..366M}. 
The significance of plunging region emission increases for lower black hole spins, which is 
attributable to the greater extent of the emission region. (see Figure 11 of \citealt{Mummery2024MNRAS.531..366M}). It has also been suggested that the plunging region may have its signature in X-ray polarization 
measurements \citep{Chan2025arXiv250415486C}.

Most IXPE observations in the soft state are well described by the standard disk model with $R_{\rm in} = R_{\rm ISCO}$. The variation of PD and PA 
with energy provides an independent method for measuring BH spins 
\citep{Dovvciak2008MNRAS.391...32D, Mikusincova2023MNRAS.519.6138M}. 
Currently, spin constraints using polarimetric data have been obtained for four sources. 
The polarimetric data for 4U~1957+115, Cygnus~X-1 and GRS~1739--278 suggest a significant role of returning radiation \citep{Marra2024A&A...684A..95M, Steiner2024ApJ...969L..30S, Zhao2026ApJ...997L..12Z}, which indicates a 
high BH spin ($a_*>0.96$ for 4U~1957+115 and Cygnus~X-1, $a_*>0.991$ for GRS~1739--278). These constraints are consistent with previous spin measurements for these three
sources using either continuum fitting (e.g., $a_*>0.9$ for 4U~1957+115, \citealt{Nowak2012ApJ...744..107N}; $a_*>0.9985$ for Cygnus~X-1, \citealt{Zhao2021ApJ...908..117Z}) or reflection methods (e.g., $a_*=0.95_{-0.04}^{+0.02}$ for 4U~1957+115, \citealt{Draghis2023ApJ...946...19D}; $0.93<a_*<0.96$ for Cygnus~X-1, \citealt{Walton2016}; $0.90<a_*<0.99$ for GRS~1739--278, \citealt{Draghis2024ApJ...969...40D}). As for LMC~X-3, only an upper limit is obtained for the BH spin from polarimetric data ($a_*<0.7$, \citealt{Svoboda2024ApJ...960....3S}), which is also consistent with previous measurements using the continuum fitting method (e.g., $a_*=0.25_{-0.29}^{+0.20}$, \citealt{Steiner2014ApJ...793L..29S}). 

However, there is an exceptional case, 4U~1630--47, where the high PD ($\sim$~8.6\%) is incompatible with the 
standard accretion disk \citep{Kushwaha2023MNRAS.524L..15K, Ratheesh2024ApJ...964...77R}. 
Several potential explanations have been proposed to account for this high PD, including the returning radiation effect, an outflowing disk atmosphere \citep{Nitindala2025A&A...694A.230N}, and electron or ion anisotropies; however, none of these explanations fully explains the observations to date \citep{Krawczynski2024ApJ...977L..10K}.


A high-energy power-law tail, often extending from 10~keV to 1~MeV without a clear cutoff, is frequently observed in the soft state spectra of black hole X-ray binaries \citep{Grove1998ApJ...500..899G, McConnell2002ApJ...572..984M}. Such a tail has also been detected in the hard and intermediate states, superimposed on the thermal Comptonization component \citep[e.g.,][]{Del-Santo2008, Zdziarski2021ApJ...914L...5Z}. In the soft state, the tail is generally attributed to Compton scattering by a non-thermal electron population (i.e., the electron energy distribution is not a Maxwellian), likely accelerated by magnetic reconnection \citep{Beloborodov2017ApJ...850..141B, Ball2018ApJ...862...80B, Sridhar2023MNRAS.518.1301S} or shocks \citep{Groselj2024PhRvL.132h5202G}. However, the spatial location of these non-thermal electrons remains unknown; they could reside in active regions above the disk \citep{Zdziarski2004} or in the plunging region \citep{Hankla2022MNRAS.515..775H}.

Although the power-law component is often weak in the soft state, it can still illuminate the disk and produce reflection features. It has been proposed that the reflection fraction (i.e., the ratio of reflected to direct coronal flux) is higher in the soft state than in the hard state \citep{Zdziarski2003MNRAS.342..355Z, Gilfanov2014SSRv..183..121G, Steiner2016ApJ...829L..22S}. This suggests that the corona is more compact in the soft state, such that light-bending effects enhance the illumination of the disk \citep{Miniutti2004MNRAS.349.1435M}. Recently, evidence has emerged indicating that returning radiation may also play a role in shaping reflection features in the soft state \citep{Connors2021, Mirzaev2024ApJ...976..229M}. This effect does not appear to significantly affect black hole spin measurements \citep{Das2025arXiv250909481D}, but its implications for our understanding of coronal geometry remain to be explored using larger source samples and more proper modeling \citep{Mirzaev2024ApJ...965...66M}.


\subsection{Soft to hard transition (Region D to A)}

The soft-to-hard state transition can occur at relatively low luminosities, e.g., 
$L_{\rm X}/L_{\rm Edd}<$ 1\% \citep{Petrucci2014}. Throughout this transition, the source evolves back through lower-luminosity variants of the soft intermediate and hard intermediate states, ultimately returning to the low-hard state.In the case of MAXI~J1820+070, the transition was monitored 
by \textit{Insight}--HXMT, which revealed that $R_{\rm in}$ gradually increased from 1 to 10 $R_{\rm ISCO}$ 
(D to E, \citealt{Fan2024ApJ...969...61F}). However, this may not be a general behavior, as observations of 
other sources remain scarce and the data quality is often lower than that of the brighter hard-to-soft transitions. 
During this same transition, time lags between the X-ray, optical, and radio bands were detected and interpreted as evidence for a magnetically arrested disk (MAD) state, in which the radial magnetic force is sufficient to balance the gravitational force \citep{You2023Sci...381..961Y}. These observed lags are thought to trace the process of magnetic field accumulation. In this scenario, the accumulation of magnetic flux near the black hole first gets strong enough to power the corona (producing X-ray emission), and subsequently becomes strong enough to launch the jet (producing the radio emission). The optical lag is attributed to a thermal-viscous instability triggered by heating of the outer disk by the X-ray flare.

A study of failed outbursts 
in GX~339--4 also found that the disk is more truncated during the rising phase (A to E) than the decaying phase 
(E to A), despite similar luminosities \citep{Wang2018, Wang2020ApJ...899...44W}. 
Such a phenomenon suggests that even within the hard state, the inner disk radius is not solely determined by the accretion rate. The accretion history of the system may also play a role. 
It is also consistent the idea that the inner disk radius is evolving within 
the hard state.
IXPE captured Swift J1727.8--1613 twice in the dim soft state (region D, \citealt{Svoboda2024ApJ...966L..35S}) and 
once during the soft-to-hard transition (near region E, \citealt{Podgorny2024A&A...686L..12P}). Compared to the bright 
hard and hard-intermediate states (PD $\sim$ 3-4\%, \citealt{Ingram2024ApJ...968...76I}), the PD drops to $<$~1\% in 
the dim soft state. At the end of the soft-to-hard transition, both the PD and PA return to the values observed 
in the bright hard state at similar hardness, despite the luminosity being over an order of 
magnitude lower \citep{Podgorny2024A&A...686L..12P}. This suggests a similar coronal 
geometry in both the bright and dim hard states.

\section{Concluding Remarks}

Over the past decades, X-ray techniques have matured and become powerful tools for studying the accretion geometry of BH XRBs. Understanding this geometry is essential for controlling systematic uncertainties when using observational data to address fundamental questions, such as testing gravity theories \citep[e.g.,][]{Liu2019, Bambi2021} or understanding black hole spin distributions \citep[e.g.,][]{Draghis2024ApJ...969...40D, Zdziarski2024ApJ...967L...9Z}. Based on a synthesis of spectral-timing results available in the literature, the following picture of the disk-corona system emerges:

\begin{enumerate}
    \item The standard accretion disk is truncated at large radii in the \textit{low hard} state, when the accretion rate is low ($L_{\rm X}/L_{\rm Edd} <$ 1\%). The radially extended hot inner flow may act as the corona, which is also consistent with X-ray polarimetric measurements.
    \item In the \textit{bright hard} state ($L_{\rm X}/L_{\rm Edd} >$ 3-4\%), the disk appears to extend close to the innermost stable circular orbit ($R_{\rm ISCO}$), before the state transition.
    \item During the hard-to-soft transition, the disk remains near $R_{\rm ISCO}$, while the corona expands vertically. X-ray polarimetry suggests that a horizontally extended coronal component may still be present \citep{Ingram2024ApJ...968...76I}. A dual-corona configuration may be needed to simultaneously interpret the X-ray spectral, timing, and polarimetric data (see Figure~\ref{geometry}). However, a self-consistent theoretical framework for this scenario is still lacking.

\end{enumerate}

IXPE observations over the past few years have significantly advanced our understanding of BH accretion geometries. 
Some results align with the picture established by traditional spectral-timing studies, while others challenge 
the traditional view \citep[e.g.,][]{Ratheesh2024ApJ...964...77R,Ewing2025MNRAS.541.1774E}. The implications of 
these challenges remain uncertain, as high-sensitivity X-ray polarimetry is still a relatively young field. 
Nevertheless, several new accretion configurations --- such as an outflowing corona \citep{Poutanen2023ApJ...949L..10P}, 
a scattering disk wind \citep{Nitindala2025A&A...694A.230N}, and a warped disk \citep{Krawczynski2022Sci...378..650K} --- have 
already been proposed based on IXPE results. Meanwhile, some older models (e.g., lamppost and spherical coronal geometries) 
have been ruled out \citep{Krawczynski2022Sci...378..650K}. An important next step is to determine which of these 
configurations most realistically explains the full suite of X-ray (and multi-wavelength) observations. 
The enhanced X-ray Timing and Polarimetry mission (eXTP), planned for launch in 2030 \citep{2016SPIE.9905E..1QZ, Zhang2025SCPMA..6819502Z, Bu2025SCPMA..6819504B}, 
will further push the field forward. eXTP will carry three X-ray polarimeters with a total effective area five 
times larger than that of IXPE, as well as a separate spectrometer capable of providing simultaneous spectral 
and timing data. Such synergistic observations are crucial for breaking degeneracies and distinguishing 
between different scenarios \citep[e.g.,][]{Ingram2024ApJ...968...76I,Mastroserio2025ApJ...978L..19M}.

The X-Ray Imaging and Spectroscopy Mission (XRISM; \citealt{Tashiro2020SPIE11444E..22T}), launched in September 2023, 
is also expected to provide new insights into BH XRBs, thanks to its high-energy-resolution microcalorimeter, 
\textit{Resolve}. This high energy resolution makes it possible to distinguish subtle spectral features and 
place tighter constraints on model parameters \citep[e.g.,][]{Garcia2022ApJ...926...13G, Liu2025MNRAS.536.2594L}. 
Some initial results from XRISM have already shown its capbility in separating the broad iron line and the narrow features \citep{Draghis2025ApJ...995L..12D, Brenneman2025ApJ...995..200B}. Fitting of the broad iron line in Cygnus~X-1 finds the inner disk inclination angle to be 30 degrees higher than the binary inclination angle, indicating a warped disk \citep{Draghis2025ApJ...995L..12D}.
The New Advanced Telescope for High-ENergy Astrophysics (NewAthena; \citealt{Cruise2025NatAs...9...36C}), 
a flagship mission planned for the late 2030s, will also carry a microcalorimeter spectrometer, 
but with a significantly larger effective area. NewAthena will enable more detailed studies of BH XRBs, as it can collect the same number of photons in just one-fourth of the exposure time required by XRISM (at 6 keV) if the gate valve shielding the
Resolve instrument can be opened in the future, or one-seventh if the gate valve remains closed.

\section{Statements and Declarations}

\noindent \textbf{Acknowledgements:} HL thanks the anonymous referee for their insightful comments. HL also thanks Ningyue Fan, Mariano Mendez, and Federico Garcia for helpful discussions, and Cosimo Bambi for proofreading the manuscript. HL acknowledges support from the IAU-Gruber Foundation Fellowship.

\noindent \textbf{Funding information:} not applicable.

\noindent \textbf{Author contributions:} H.L. prepared this manuscript as a single author.

\noindent \textbf{Ethics declaration:} not applicable.

\noindent \textbf{Data availability:} No datasets were generated or analyzed.

\bibliography{sn-bibliography}

@ARTICLE{Jiang2020MNRAS.492.1947J,
       author = {{Jiang}, Jiachen and {F{\"u}rst}, Felix and {Walton}, Dominic J. and {Parker}, Michael L. and {Fabian}, Andrew C.},
        title = "{A NuSTAR view of GRS 1716-249 in the hard and intermediate states}",
      journal = {\mnras},
         year = 2020,
        month = feb,
       volume = {492},
       number = {2},
        pages = {1947-1956},
          doi = {10.1093/mnras/staa017},
archivePrefix = {arXiv},
       eprint = {1912.13215},
 primaryClass = {astro-ph.HE},
       adsurl = {https://ui.adsabs.harvard.edu/abs/2020MNRAS.492.1947J}
}

@ARTICLE{Parker2016,
       author = {{Parker}, M.~L. and {Tomsick}, J.~A. and {Kennea}, J.~A. and {Miller}, J.~M. and {Harrison}, F.~A. and {Barret}, D. and {Boggs}, S.~E. and {Christensen}, F.~E. and {Craig}, W.~W. and {Fabian}, A.~C. and {F{\"u}rst}, F. and {Grinberg}, V. and {Hailey}, C.~J. and {Romano}, P. and {Stern}, D. and {Walton}, D.~J. and {Zhang}, W.~W.},
        title = "{NuSTAR and Swift Observations of the Very High State in GX 339-4: Weighing the Black Hole with X-Rays}",
      journal = {\apjl},
         year = 2016,
        month = apr,
       volume = {821},
       number = {1},
          eid = {L6},
        pages = {L6},
          doi = {10.3847/2041-8205/821/1/L6},
archivePrefix = {arXiv},
       eprint = {1603.03777},
 primaryClass = {astro-ph.HE},
       adsurl = {https://ui.adsabs.harvard.edu/abs/2016ApJ...821L...6P}
}

@ARTICLE{Jiang2019gx339,
       author = {{Jiang}, Jiachen and {Fabian}, Andrew C. and {Wang}, Jingyi and {Walton}, Dominic J. and {Garc{\'\i}a}, Javier A. and {Parker}, Michael L. and {Steiner}, James F. and {Tomsick}, John A.},
        title = "{High-density reflection spectroscopy: I. A case study of GX 339-4}",
      journal = {\mnras},
         year = 2019,
        month = apr,
       volume = {484},
       number = {2},
        pages = {1972-1982},
          doi = {10.1093/mnras/stz095},
archivePrefix = {arXiv},
       eprint = {1901.01739},
 primaryClass = {astro-ph.HE},
       adsurl = {https://ui.adsabs.harvard.edu/abs/2019MNRAS.484.1972J}
}

@ARTICLE{Garcia2015,
       author = {{Garc{\'\i}a}, Javier A. and {Steiner}, James F. and {McClintock}, Jeffrey E. and {Remillard}, Ronald A. and {Grinberg}, Victoria and {Dauser}, Thomas},
        title = "{X-Ray Reflection Spectroscopy of the Black Hole GX 339--4: Exploring the Hard State with Unprecedented Sensitivity}",
      journal = {\apj},
         year = 2015,
        month = nov,
       volume = {813},
       number = {2},
          eid = {84},
        pages = {84},
          doi = {10.1088/0004-637X/813/2/84},
archivePrefix = {arXiv},
       eprint = {1505.03607},
 primaryClass = {astro-ph.HE},
       adsurl = {https://ui.adsabs.harvard.edu/abs/2015ApJ...813...84G}
}

@ARTICLE{Belloni2005,
       author = {{Belloni}, T. and {Homan}, J. and {Casella}, P. and {van der Klis}, M. and {Nespoli}, E. and {Lewin}, W.~H.~G. and {Miller}, J.~M. and {M{\'e}ndez}, M.},
        title = "{The evolution of the timing properties of the black-hole transient GX 339-4 during its 2002/2003 outburst}",
      journal = {\aap},
         year = 2005,
        month = sep,
       volume = {440},
       number = {1},
        pages = {207-222},
          doi = {10.1051/0004-6361:20042457},
archivePrefix = {arXiv},
       eprint = {astro-ph/0504577},
 primaryClass = {astro-ph},
       adsurl = {https://ui.adsabs.harvard.edu/abs/2005A&A...440..207B}
}

@ARTICLE{Remillard2006,
       author = {{Remillard}, Ronald A. and {McClintock}, Jeffrey E.},
        title = "{X-Ray Properties of Black-Hole Binaries}",
      journal = {\araa},
         year = 2006,
        month = sep,
       volume = {44},
       number = {1},
        pages = {49-92},
          doi = {10.1146/annurev.astro.44.051905.092532},
archivePrefix = {arXiv},
       eprint = {astro-ph/0606352},
 primaryClass = {astro-ph},
       adsurl = {https://ui.adsabs.harvard.edu/abs/2006ARA&A..44...49R}
}

@ARTICLE{Fender2004,
       author = {{Fender}, R.~P. and {Belloni}, T.~M. and {Gallo}, E.},
        title = "{Towards a unified model for black hole X-ray binary jets}",
      journal = {\mnras},
         year = 2004,
        month = dec,
       volume = {355},
       number = {4},
        pages = {1105-1118},
          doi = {10.1111/j.1365-2966.2004.08384.x},
archivePrefix = {arXiv},
       eprint = {astro-ph/0409360},
 primaryClass = {astro-ph},
       adsurl = {https://ui.adsabs.harvard.edu/abs/2004MNRAS.355.1105F}
}

@ARTICLE{Done2007,
       author = {{Done}, Chris and {Gierli{\'n}ski}, Marek and {Kubota}, Aya},
        title = "{Modelling the behaviour of accretion flows in X-ray binaries. Everything you always wanted to know about accretion but were afraid to ask}",
      journal = {\aapr},
         year = 2007,
        month = dec,
       volume = {15},
       number = {1},
        pages = {1-66},
          doi = {10.1007/s00159-007-0006-1},
archivePrefix = {arXiv},
       eprint = {0708.0148},
 primaryClass = {astro-ph},
       adsurl = {https://ui.adsabs.harvard.edu/abs/2007A&ARv..15....1D}
}

@ARTICLE{Zdziarski2019,
       author = {{Zdziarski}, Andrzej A. and {Zi{\'o}{\l}kowski}, Janusz and {Miko{\l}ajewska}, Joanna},
        title = "{The X-ray binary GX 339-4/V821 Ara: the distance, inclination, evolutionary status, and mass transfer}",
      journal = {\mnras},
         year = 2019,
        month = sep,
       volume = {488},
       number = {1},
        pages = {1026-1034},
          doi = {10.1093/mnras/stz1787},
archivePrefix = {arXiv},
       eprint = {1904.07803},
 primaryClass = {astro-ph.SR},
       adsurl = {https://ui.adsabs.harvard.edu/abs/2019MNRAS.488.1026Z}
}

@ARTICLE{Dauser2010,
       author = {{Dauser}, T. and {Wilms}, J. and {Reynolds}, C.~S. and {Brenneman}, L.~W.},
        title = "{Broad emission lines for a negatively spinning black hole}",
      journal = {\mnras},
         year = 2010,
        month = dec,
       volume = {409},
       number = {4},
        pages = {1534-1540},
          doi = {10.1111/j.1365-2966.2010.17393.x},
archivePrefix = {arXiv},
       eprint = {1007.4937},
 primaryClass = {astro-ph.HE},
       adsurl = {https://ui.adsabs.harvard.edu/abs/2010MNRAS.409.1534D}
}

@ARTICLE{Mitsuda1984,
       author = {{Mitsuda}, K. and {Inoue}, H. and {Koyama}, K. and {Makishima}, K. and {Matsuoka}, M. and {Ogawara}, Y. and {Shibazaki}, N. and {Suzuki}, K. and {Tanaka}, Y. and {Hirano}, T.},
        title = "{Energy spectra of low-mass binary X-ray sources observed from Tenma.}",
      journal = {\pasj},
         year = 1984,
        month = jan,
       volume = {36},
        pages = {741-759},
       adsurl = {https://ui.adsabs.harvard.edu/abs/1984PASJ...36..741M}
}

@ARTICLE{Zdziarski2004,
       author = {{Zdziarski}, A.~A. and {Gierli{\'n}ski}, M.},
        title = "{Radiative Processes, Spectral States and Variability of Black-Hole Binaries}",
      journal = {Progress of Theoretical Physics Supplement},
         year = 2004,
        month = jan,
       volume = {155},
        pages = {99-119},
          doi = {10.1143/PTPS.155.99},
archivePrefix = {arXiv},
       eprint = {astro-ph/0403683},
 primaryClass = {astro-ph},
       adsurl = {https://ui.adsabs.harvard.edu/abs/2004PThPS.155...99Z}
}

@ARTICLE{Bambi2021,
       author = {{Bambi}, Cosimo and {Brenneman}, Laura W. and {Dauser}, Thomas and {Garc{\'\i}a}, Javier A. and {Grinberg}, Victoria and {Ingram}, Adam and {Jiang}, Jiachen and {Liu}, Honghui and {Lohfink}, Anne M. and {Marinucci}, Andrea and {Mastroserio}, Guglielmo and {Middei}, Riccardo and {Nampalliwar}, Sourabh and {Nied{\'z}wiecki}, Andrzej and {Steiner}, James F. and {Tripathi}, Ashutosh and {Zdziarski}, Andrzej A.},
        title = "{Towards Precision Measurements of Accreting Black Holes Using X-Ray Reflection Spectroscopy}",
      journal = {\ssr},
         year = 2021,
        month = aug,
       volume = {217},
       number = {5},
          eid = {65},
        pages = {65},
          doi = {10.1007/s11214-021-00841-8},
archivePrefix = {arXiv},
       eprint = {2011.04792},
 primaryClass = {astro-ph.HE},
       adsurl = {https://ui.adsabs.harvard.edu/abs/2021SSRv..217...65B}
}

@ARTICLE{Bambi2017,
       author = {{Bambi}, Cosimo},
        title = "{Testing black hole candidates with electromagnetic radiation}",
      journal = {Reviews of Modern Physics},
         year = 2017,
        month = apr,
       volume = {89},
       number = {2},
          eid = {025001},
        pages = {025001},
          doi = {10.1103/RevModPhys.89.025001},
archivePrefix = {arXiv},
       eprint = {1509.03884},
 primaryClass = {gr-qc},
       adsurl = {https://ui.adsabs.harvard.edu/abs/2017RvMP...89b5001B}
}

@ARTICLE{Liu2019,
       author = {{Liu}, Honghui and {Abdikamalov}, Askar B. and {Ayzenberg}, Dimitry and {Bambi}, Cosimo and {Dauser}, Thomas and {Garc{\'\i}a}, Javier A. and {Nampalliwar}, Sourabh},
        title = "{Testing the Kerr hypothesis using x-ray reflection spectroscopy with NuSTAR data of Cygnus X-1 in the soft state}",
      journal = {\prd},
         year = 2019,
        month = jun,
       volume = {99},
       number = {12},
          eid = {123007},
        pages = {123007},
          doi = {10.1103/PhysRevD.99.123007},
archivePrefix = {arXiv},
       eprint = {1904.08027},
 primaryClass = {gr-qc},
       adsurl = {https://ui.adsabs.harvard.edu/abs/2019PhRvD..99l3007L}
}

@ARTICLE{Furst2015,
       author = {{F{\"u}rst}, F. and {Nowak}, M.~A. and {Tomsick}, J.~A. and {Miller}, J.~M. and {Corbel}, S. and {Bachetti}, M. and {Boggs}, S.~E. and {Christensen}, F.~E. and {Craig}, W.~W. and {Fabian}, A.~C. and {Gandhi}, P. and {Grinberg}, V. and {Hailey}, C.~J. and {Harrison}, F.~A. and {Kara}, E. and {Kennea}, J.~A. and {Madsen}, K.~K. and {Pottschmidt}, K. and {Stern}, D. and {Walton}, D.~J. and {Wilms}, J. and {Zhang}, W.~W.},
        title = "{The Complex Accretion Geometry of GX 339-4 as Seen by NuSTAR and Swift}",
      journal = {\apj},
         year = 2015,
        month = aug,
       volume = {808},
       number = {2},
          eid = {122},
        pages = {122},
          doi = {10.1088/0004-637X/808/2/122},
archivePrefix = {arXiv},
       eprint = {1506.01381},
 primaryClass = {astro-ph.HE},
       adsurl = {https://ui.adsabs.harvard.edu/abs/2015ApJ...808..122F}
}

@ARTICLE{Homan2005,
       author = {{Homan}, Jeroen and {Belloni}, Tomaso},
        title = "{The Evolution of Black Hole States}",
      journal = {\apss},
         year = 2005,
        month = nov,
       volume = {300},
       number = {1-3},
        pages = {107-117},
          doi = {10.1007/s10509-005-1197-4},
archivePrefix = {arXiv},
       eprint = {astro-ph/0412597},
 primaryClass = {astro-ph},
       adsurl = {https://ui.adsabs.harvard.edu/abs/2005Ap&SS.300..107H}
}

@ARTICLE{Fabian1989,
       author = {{Fabian}, A.~C. and {Rees}, M.~J. and {Stella}, L. and {White}, N.~E.},
        title = "{X-ray fluorescence from the inner disc in Cygnus X-1.}",
      journal = {\mnras},
         year = 1989,
        month = may,
       volume = {238},
        pages = {729-736},
          doi = {10.1093/mnras/238.3.729},
       adsurl = {https://ui.adsabs.harvard.edu/abs/1989MNRAS.238..729F}
}

@ARTICLE{Fabian2000,
       author = {{Fabian}, A.~C. and {Iwasawa}, K. and {Reynolds}, C.~S. and {Young}, A.~J.},
        title = "{Broad Iron Lines in Active Galactic Nuclei}",
      journal = {\pasp},
         year = 2000,
        month = sep,
       volume = {112},
       number = {775},
        pages = {1145-1161},
          doi = {10.1086/316610},
archivePrefix = {arXiv},
       eprint = {astro-ph/0004366},
 primaryClass = {astro-ph},
       adsurl = {https://ui.adsabs.harvard.edu/abs/2000PASP..112.1145F}
}

@ARTICLE{Esin1997,
       author = {{Esin}, Ann A. and {McClintock}, Jeffrey E. and {Narayan}, Ramesh},
        title = "{Advection-Dominated Accretion and the Spectral States of Black Hole X-Ray Binaries: Application to Nova Muscae 1991}",
      journal = {\apj},
         year = 1997,
        month = nov,
       volume = {489},
       number = {2},
        pages = {865-889},
          doi = {10.1086/304829},
archivePrefix = {arXiv},
       eprint = {astro-ph/9705237},
 primaryClass = {astro-ph},
       adsurl = {https://ui.adsabs.harvard.edu/abs/1997ApJ...489..865E}
}

@ARTICLE{Reis2008,
       author = {{Reis}, R.~C. and {Fabian}, A.~C. and {Ross}, R.~R. and {Miniutti}, G. and {Miller}, J.~M. and {Reynolds}, C.},
        title = "{A systematic look at the very high and low/hard state of GX339-4: constraining the black hole spin with a new reflection model}",
      journal = {\mnras},
         year = 2008,
        month = jul,
       volume = {387},
       number = {4},
        pages = {1489-1498},
          doi = {10.1111/j.1365-2966.2008.13358.x},
archivePrefix = {arXiv},
       eprint = {0804.0238},
 primaryClass = {astro-ph},
       adsurl = {https://ui.adsabs.harvard.edu/abs/2008MNRAS.387.1489R}
}

@ARTICLE{Reis2009,
       author = {{Reis}, R.~C. and {Fabian}, A.~C. and {Ross}, R.~R. and {Miller}, J.~M.},
        title = "{Determining the spin of two stellar-mass black holes from disc reflection signatures}",
      journal = {\mnras},
         year = 2009,
        month = may,
       volume = {395},
       number = {3},
        pages = {1257-1264},
          doi = {10.1111/j.1365-2966.2009.14622.x},
       adsurl = {https://ui.adsabs.harvard.edu/abs/2009MNRAS.395.1257R}
}

@ARTICLE{Reis2011,
       author = {{Reis}, R.~C. and {Miller}, J.~M. and {Fabian}, A.~C. and {Cackett}, E.~M. and {Maitra}, D. and {Reynolds}, C.~S. and {Rupen}, M. and {Steeghs}, D.~T.~H. and {Wijnands}, R.},
        title = "{Multistate observations of the Galactic black hole XTE J1752-223: evidence for an intermediate black hole spin}",
      journal = {\mnras},
         year = 2011,
        month = feb,
       volume = {410},
       number = {4},
        pages = {2497-2505},
          doi = {10.1111/j.1365-2966.2010.17628.x},
archivePrefix = {arXiv},
       eprint = {1009.1154},
 primaryClass = {astro-ph.HE},
       adsurl = {https://ui.adsabs.harvard.edu/abs/2011MNRAS.410.2497R}
}

@ARTICLE{Tomsick2008ApJ...680..593T,
       author = {{Tomsick}, John A. and {Kalemci}, Emrah and {Kaaret}, Philip and {Markoff}, Sera and {Corbel}, Stephane and {Migliari}, Simone and {Fender}, Rob and {Bailyn}, Charles D. and {Buxton}, Michelle M.},
        title = "{Broadband X-Ray Spectra of GX 339-4 and the Geometry of Accreting Black Holes in the Hard State}",
      journal = {\apj},
         year = 2008,
        month = jun,
       volume = {680},
       number = {1},
        pages = {593-601},
          doi = {10.1086/587797},
archivePrefix = {arXiv},
       eprint = {0802.3357},
 primaryClass = {astro-ph},
       adsurl = {https://ui.adsabs.harvard.edu/abs/2008ApJ...680..593T}
}

@INCOLLECTION{Belloni2010,
       author = {{Belloni}, T.~M.},
        title = "{States and Transitions in Black Hole Binaries}",
    booktitle = {Lecture Notes in Physics, Berlin Springer Verlag},
         year = 2010,
       editor = {{Belloni}, Tomaso},
       volume = {794},
        pages = {53},
          doi = {10.1007/978-3-540-76937-8\_3},
       adsurl = {https://ui.adsabs.harvard.edu/abs/2010LNP...794...53B}
}

@ARTICLE{Dauser2013,
       author = {{Dauser}, T. and {Garcia}, J. and {Wilms}, J. and {B{\"o}ck}, M. and {Brenneman}, L.~W. and {Falanga}, M. and {Fukumura}, K. and {Reynolds}, C.~S.},
        title = "{Irradiation of an accretion disc by a jet: general properties and implications for spin measurements of black holes}",
      journal = {\mnras},
         year = 2013,
        month = apr,
       volume = {430},
       number = {3},
        pages = {1694-1708},
          doi = {10.1093/mnras/sts710},
archivePrefix = {arXiv},
       eprint = {1301.4922},
 primaryClass = {astro-ph.HE},
       adsurl = {https://ui.adsabs.harvard.edu/abs/2013MNRAS.430.1694D}
}

@ARTICLE{Wang2018,
       author = {{Wang-Ji}, Jingyi and {Garc{\'\i}a}, Javier A. and {Steiner}, James F. and {Tomsick}, John A. and {Harrison}, Fiona A. and {Bambi}, Cosimo and {Petrucci}, Pierre-Olivier and {Ferreira}, Jonathan and {Chakravorty}, Susmita and {Clavel}, Ma{\"\i}ca},
        title = "{The Evolution of GX 339-4 in the Low-hard State as Seen by NuSTAR and Swift}",
      journal = {\apj},
         year = 2018,
        month = mar,
       volume = {855},
       number = {1},
          eid = {61},
        pages = {61},
          doi = {10.3847/1538-4357/aaa974},
archivePrefix = {arXiv},
       eprint = {1712.02571},
 primaryClass = {astro-ph.HE},
       adsurl = {https://ui.adsabs.harvard.edu/abs/2018ApJ...855...61W}
}

@ARTICLE{Sridhar2020,
       author = {{Sridhar}, Navin and {Garc{\'\i}a}, Javier A. and {Steiner}, James F. and {Connors}, Riley M.~T. and {Grinberg}, Victoria and {Harrison}, Fiona A.},
        title = "{Evolution of the Accretion Disk-Corona during the Bright Hard-to-soft State Transition: A Reflection Spectroscopic Study with GX 339-4}",
      journal = {\apj},
         year = 2020,
        month = feb,
       volume = {890},
       number = {1},
          eid = {53},
        pages = {53},
          doi = {10.3847/1538-4357/ab64f5},
archivePrefix = {arXiv},
       eprint = {1912.11447},
 primaryClass = {astro-ph.HE},
       adsurl = {https://ui.adsabs.harvard.edu/abs/2020ApJ...890...53S}
}

@ARTICLE{Garcia2010,
       author = {{Garc{\'\i}a}, J. and {Kallman}, T.~R.},
        title = "{X-ray Reflected Spectra from Accretion Disk Models. I. Constant Density Atmospheres}",
      journal = {\apj},
         year = 2010,
        month = aug,
       volume = {718},
       number = {2},
        pages = {695-706},
          doi = {10.1088/0004-637X/718/2/695},
archivePrefix = {arXiv},
       eprint = {1006.0485},
 primaryClass = {astro-ph.HE},
       adsurl = {https://ui.adsabs.harvard.edu/abs/2010ApJ...718..695G}
}

@ARTICLE{Petrucci2014,
       author = {{Petrucci}, P. -O. and {Cabanac}, C. and {Corbel}, S. and {Koerding}, E. and {Fender}, R.},
        title = "{The return to the hard state of GX 339-4 as seen by Suzaku}",
      journal = {\aap},
         year = 2014,
        month = apr,
       volume = {564},
          eid = {A37},
        pages = {A37},
          doi = {10.1051/0004-6361/201322268},
archivePrefix = {arXiv},
       eprint = {1310.3039},
 primaryClass = {astro-ph.HE},
       adsurl = {https://ui.adsabs.harvard.edu/abs/2014A&A...564A..37P}
}

@ARTICLE{Shidatsu2011,
       author = {{Shidatsu}, Megumi and {Ueda}, Yoshihiro and {Tazaki}, Fumie and {Yoshikawa}, Tatsuhito and {Nagayama}, Takahiro and {Nagata}, Tetsuya and {Oi}, Nagisa and {Yamaoka}, Kazutaka and {Takahashi}, Hiromitsu and {Kubota}, Aya and {Cottam}, Jean and {Remillard}, Ronald and {Negoro}, Hitoshi},
        title = "{X-Ray and Near-Infrared Observations of GX 339-4 in the Low/Hard State with Suzaku and IRSF}",
      journal = {\pasj},
         year = 2011,
        month = nov,
       volume = {63},
        pages = {S785-S801},
          doi = {10.1093/pasj/63.sp3.S785},
archivePrefix = {arXiv},
       eprint = {1105.3586},
 primaryClass = {astro-ph.HE},
       adsurl = {https://ui.adsabs.harvard.edu/abs/2011PASJ...63S.785S}
}

@ARTICLE{Tomsick2009,
       author = {{Tomsick}, John A. and {Yamaoka}, Kazutaka and {Corbel}, Stephane and {Kaaret}, Philip and {Kalemci}, Emrah and {Migliari}, Simone},
        title = "{Truncation of the Inner Accretion Disk Around a Black Hole at Low Luminosity}",
      journal = {\apjl},
         year = 2009,
        month = dec,
       volume = {707},
       number = {1},
        pages = {L87-L91},
          doi = {10.1088/0004-637X/707/1/L87},
archivePrefix = {arXiv},
       eprint = {0911.2240},
 primaryClass = {astro-ph.HE},
       adsurl = {https://ui.adsabs.harvard.edu/abs/2009ApJ...707L..87T}
}

@ARTICLE{Steiner2010,
       author = {{Steiner}, James F. and {McClintock}, Jeffrey E. and {Remillard}, Ronald A. and {Gou}, Lijun and {Yamada}, Shin'ya and {Narayan}, Ramesh},
        title = "{The Constant Inner-disk Radius of LMC X-3: A Basis for Measuring Black Hole Spin}",
      journal = {\apjl},
         year = 2010,
        month = aug,
       volume = {718},
       number = {2},
        pages = {L117-L121},
          doi = {10.1088/2041-8205/718/2/L117},
archivePrefix = {arXiv},
       eprint = {1006.5729},
 primaryClass = {astro-ph.HE},
       adsurl = {https://ui.adsabs.harvard.edu/abs/2010ApJ...718L.117S}
}

@ARTICLE{Gierlinski2004,
       author = {{Gierli{\'n}ski}, Marek and {Done}, Chris},
        title = "{Black hole accretion discs: reality confronts theory}",
      journal = {\mnras},
         year = 2004,
        month = jan,
       volume = {347},
       number = {3},
        pages = {885-894},
          doi = {10.1111/j.1365-2966.2004.07266.x},
archivePrefix = {arXiv},
       eprint = {astro-ph/0307333},
 primaryClass = {astro-ph},
       adsurl = {https://ui.adsabs.harvard.edu/abs/2004MNRAS.347..885G}
}

@ARTICLE{Del-Santo2008,
       author = {{Del Santo}, M. and {Malzac}, J. and {Jourdain}, E. and {Belloni}, T. and {Ubertini}, P.},
        title = "{Spectral variability of GX339-4 in a hard-to-soft state transition}",
      journal = {\mnras},
         year = 2008,
        month = oct,
       volume = {390},
       number = {1},
        pages = {227-234},
          doi = {10.1111/j.1365-2966.2008.13672.x},
archivePrefix = {arXiv},
       eprint = {0807.1018},
 primaryClass = {astro-ph},
       adsurl = {https://ui.adsabs.harvard.edu/abs/2008MNRAS.390..227D}
}

@ARTICLE{Wang2021,
       author = {{Wang}, Jingyi and {Mastroserio}, Guglielmo and {Kara}, Erin and {Garc{\'\i}a}, Javier A. and {Ingram}, Adam and {Connors}, Riley and {van der Klis}, Michiel and {Dauser}, Thomas and {Steiner}, James F. and {Buisson}, Douglas J.~K. and {Homan}, Jeroen and {Lucchini}, Matteo and {Fabian}, Andrew C. and {Bright}, Joe and {Fender}, Rob and {Cackett}, Edward M. and {Remillard}, Ron A.},
        title = "{Disk, Corona, Jet Connection in the Intermediate State of MAXI J1820+070 Revealed by NICER Spectral-timing Analysis}",
      journal = {\apjl},
         year = 2021,
        month = mar,
       volume = {910},
       number = {1},
          eid = {L3},
        pages = {L3},
          doi = {10.3847/2041-8213/abec79},
archivePrefix = {arXiv},
       eprint = {2103.05616},
 primaryClass = {astro-ph.HE},
       adsurl = {https://ui.adsabs.harvard.edu/abs/2021ApJ...910L...3W}
}

@ARTICLE{Buisson2019,
       author = {{Buisson}, D.~J.~K. and {Fabian}, A.~C. and {Barret}, D. and {F{\"u}rst}, F. and {Gandhi}, P. and {Garc{\'\i}a}, J.~A. and {Kara}, E. and {Madsen}, K.~K. and {Miller}, J.~M. and {Parker}, M.~L. and {Shaw}, A.~W. and {Tomsick}, J.~A. and {Walton}, D.~J.},
        title = "{MAXI J1820+070 with NuSTAR I. An increase in variability frequency but a stable reflection spectrum: coronal properties and implications for the inner disc in black hole binaries}",
      journal = {\mnras},
         year = 2019,
        month = nov,
       volume = {490},
       number = {1},
        pages = {1350-1362},
          doi = {10.1093/mnras/stz2681},
archivePrefix = {arXiv},
       eprint = {1909.04688},
 primaryClass = {astro-ph.HE},
       adsurl = {https://ui.adsabs.harvard.edu/abs/2019MNRAS.490.1350B}
}

@ARTICLE{George1991,
       author = {{George}, I.~M. and {Fabian}, A.~C.},
        title = "{X-ray reflection from cold matter in Active Galactic Nuclei and X-ray binaries.}",
      journal = {\mnras},
         year = 1991,
        month = mar,
       volume = {249},
        pages = {352},
          doi = {10.1093/mnras/249.2.352},
       adsurl = {https://ui.adsabs.harvard.edu/abs/1991MNRAS.249..352G}
}

@ARTICLE{Connors2021,
       author = {{Connors}, Riley M.~T. and {Garc{\'\i}a}, Javier A. and {Tomsick}, John and {Hare}, Jeremy and {Dauser}, Thomas and {Grinberg}, Victoria and {Steiner}, James F. and {Mastroserio}, Guglielmo and {Sridhar}, Navin and {Fabian}, Andrew C. and {Jiang}, Jiachen and {Parker}, Michael L. and {Harrison}, Fiona and {Kallman}, Timothy R.},
        title = "{Reflection Modeling of the Black Hole Binary 4U 1630-47: The Disk Density and Returning Radiation}",
      journal = {\apj},
         year = 2021,
        month = mar,
       volume = {909},
       number = {2},
          eid = {146},
        pages = {146},
          doi = {10.3847/1538-4357/abdd2c},
archivePrefix = {arXiv},
       eprint = {2101.06343},
 primaryClass = {astro-ph.HE},
       adsurl = {https://ui.adsabs.harvard.edu/abs/2021ApJ...909..146C}
}

@ARTICLE{Ross2007,
       author = {{Ross}, R.~R. and {Fabian}, A.~C.},
        title = "{X-ray reflection in accreting stellar-mass black hole systems}",
      journal = {\mnras},
         year = 2007,
        month = nov,
       volume = {381},
       number = {4},
        pages = {1697-1701},
          doi = {10.1111/j.1365-2966.2007.12339.x},
archivePrefix = {arXiv},
       eprint = {0709.0270},
 primaryClass = {astro-ph},
       adsurl = {https://ui.adsabs.harvard.edu/abs/2007MNRAS.381.1697R}
}

@ARTICLE{Wang2022ApJ...930...18W,
       author = {{Wang}, Jingyi and {Kara}, Erin and {Lucchini}, Matteo and {Ingram}, Adam and {van der Klis}, Michiel and {Mastroserio}, Guglielmo and {Garc{\'\i}a}, Javier A. and {Dauser}, Thomas and {Connors}, Riley and {Fabian}, Andrew C. and {Steiner}, James F. and {Remillard}, Ron A. and {Cackett}, Edward M. and {Uttley}, Phil and {Altamirano}, Diego},
        title = "{The NICER ``Reverberation Machine'': A Systematic Study of Time Lags in Black Hole X-Ray Binaries}",
      journal = {\apj},
         year = 2022,
        month = may,
       volume = {930},
       number = {1},
          eid = {18},
        pages = {18},
          doi = {10.3847/1538-4357/ac6262},
archivePrefix = {arXiv},
       eprint = {2205.00928},
 primaryClass = {astro-ph.HE},
       adsurl = {https://ui.adsabs.harvard.edu/abs/2022ApJ...930...18W}
}

@ARTICLE{Ross2005,
       author = {{Ross}, R.~R. and {Fabian}, A.~C.},
        title = "{A comprehensive range of X-ray ionized-reflection models}",
      journal = {\mnras},
         year = 2005,
        month = mar,
       volume = {358},
       number = {1},
        pages = {211-216},
          doi = {10.1111/j.1365-2966.2005.08797.x},
archivePrefix = {arXiv},
       eprint = {astro-ph/0501116},
 primaryClass = {astro-ph},
       adsurl = {https://ui.adsabs.harvard.edu/abs/2005MNRAS.358..211R}
}

@ARTICLE{Steiner2012,
       author = {{Steiner}, James F. and {McClintock}, Jeffrey E. and {Reid}, Mark J.},
        title = "{The Distance, Inclination, and Spin of the Black Hole Microquasar H1743-322}",
      journal = {\apjl},
         year = 2012,
        month = jan,
       volume = {745},
       number = {1},
          eid = {L7},
        pages = {L7},
          doi = {10.1088/2041-8205/745/1/L7},
archivePrefix = {arXiv},
       eprint = {1111.2388},
 primaryClass = {astro-ph.HE},
       adsurl = {https://ui.adsabs.harvard.edu/abs/2012ApJ...745L...7S}
}

@ARTICLE{Walton2017,
       author = {{Walton}, D.~J. and {Mooley}, K. and {King}, A.~L. and {Tomsick}, J.~A. and {Miller}, J.~M. and {Dauser}, T. and {Garc{\'\i}a}, J.~A. and {Bachetti}, M. and {Brightman}, M. and {Fabian}, A.~C. and {Forster}, K. and {F{\"u}rst}, F. and {Gandhi}, P. and {Grefenstette}, B.~W. and {Harrison}, F.~A. and {Madsen}, K.~K. and {Meier}, D.~L. and {Middleton}, M.~J. and {Natalucci}, L. and {Rahoui}, F. and {Rana}, V. and {Stern}, D.},
        title = "{Living on a Flare: Relativistic Reflection in V404 Cyg Observed by NuSTAR during Its Summer 2015 Outburst}",
      journal = {\apj},
         year = 2017,
        month = apr,
       volume = {839},
       number = {2},
          eid = {110},
        pages = {110},
          doi = {10.3847/1538-4357/aa67e8},
archivePrefix = {arXiv},
       eprint = {1609.01293},
 primaryClass = {astro-ph.HE},
       adsurl = {https://ui.adsabs.harvard.edu/abs/2017ApJ...839..110W}
}

@ARTICLE{Rodriguez2011,
       author = {{Rodriguez}, J. and {Corbel}, S. and {Caballero}, I. and {Tomsick}, J.~A. and {Tzioumis}, T. and {Paizis}, A. and {Cadolle Bel}, M. and {Kuulkers}, E.},
        title = "{First simultaneous multi-wavelength observations of the black hole candidate IGR J17091-3624. ATCA, INTEGRAL, Swift, and RXTE views of the 2011 outburst}",
      journal = {\aap},
         year = 2011,
        month = sep,
       volume = {533},
          eid = {L4},
        pages = {L4},
          doi = {10.1051/0004-6361/201117511},
archivePrefix = {arXiv},
       eprint = {1108.0666},
 primaryClass = {astro-ph.HE},
       adsurl = {https://ui.adsabs.harvard.edu/abs/2011A&A...533L...4R}
}

@ARTICLE{Reis2012ApJ...751...34R,
       author = {{Reis}, R.~C. and {Miller}, J.~M. and {Reynolds}, M.~T. and {Fabian}, A.~C. and {Walton}, D.~J.},
        title = "{Suzaku Observation of the Black Hole Candidate Maxi J1836-194 in a Hard/Intermediate Spectral State}",
      journal = {\apj},
         year = 2012,
        month = may,
       volume = {751},
       number = {1},
          eid = {34},
        pages = {34},
          doi = {10.1088/0004-637X/751/1/34},
archivePrefix = {arXiv},
       eprint = {1111.6665},
 primaryClass = {astro-ph.HE},
       adsurl = {https://ui.adsabs.harvard.edu/abs/2012ApJ...751...34R}
}

@ARTICLE{Steiner2012MNRAS.427.2552S,
       author = {{Steiner}, James F. and {Reis}, Rubens C. and {Fabian}, Andrew C. and {Remillard}, Ronald A. and {McClintock}, Jeffrey E. and {Gou}, Lijun and {Cooke}, Ryan and {Brenneman}, Laura W. and {Sanders}, Jeremy S.},
        title = "{A broad iron line in LMC X-1}",
      journal = {\mnras},
         year = 2012,
        month = dec,
       volume = {427},
       number = {3},
        pages = {2552-2561},
          doi = {10.1111/j.1365-2966.2012.22128.x},
archivePrefix = {arXiv},
       eprint = {1209.3269},
 primaryClass = {astro-ph.HE},
       adsurl = {https://ui.adsabs.harvard.edu/abs/2012MNRAS.427.2552S}
}

@ARTICLE{Walton2012MNRAS.422.2510W,
       author = {{Walton}, D.~J. and {Reis}, R.~C. and {Cackett}, E.~M. and {Fabian}, A.~C. and {Miller}, J.~M.},
        title = "{The similarity of broad iron lines in X-ray binaries and active galactic nuclei}",
      journal = {\mnras},
         year = 2012,
        month = may,
       volume = {422},
       number = {3},
        pages = {2510-2531},
          doi = {10.1111/j.1365-2966.2012.20809.x},
archivePrefix = {arXiv},
       eprint = {1202.5193},
 primaryClass = {astro-ph.HE},
       adsurl = {https://ui.adsabs.harvard.edu/abs/2012MNRAS.422.2510W}
}

@ARTICLE{King2014ApJ...784L...2K,
       author = {{King}, Ashley L. and {Walton}, Dominic J. and {Miller}, Jon M. and {Barret}, Didier and {Boggs}, Steven E. and {Christensen}, Finn E. and {Craig}, William W. and {Fabian}, Andy C. and {F{\"u}rst}, Felix and {Hailey}, Charles J. and {Harrison}, Fiona A. and {Krivonos}, Roman and {Mori}, Kaya and {Natalucci}, Lorenzo and {Stern}, Daniel and {Tomsick}, John A. and {Zhang}, William W.},
        title = "{The Disk Wind in the Rapidly Spinning Stellar-mass Black Hole 4U 1630-472 Observed with NuSTAR}",
      journal = {\apjl},
         year = 2014,
        month = mar,
       volume = {784},
       number = {1},
          eid = {L2},
        pages = {L2},
          doi = {10.1088/2041-8205/784/1/L2},
archivePrefix = {arXiv},
       eprint = {1401.3646},
 primaryClass = {astro-ph.HE},
       adsurl = {https://ui.adsabs.harvard.edu/abs/2014ApJ...784L...2K}
}

@ARTICLE{Chiang2012MNRAS.425.2436C,
       author = {{Chiang}, Chia-Ying and {Reis}, R.~C. and {Walton}, D.~J. and {Fabian}, A.~C.},
        title = "{Re-examining the XMM-Newton spectrum of the black hole candidate XTE J1652-453}",
      journal = {\mnras},
         year = 2012,
        month = oct,
       volume = {425},
       number = {4},
        pages = {2436-2442},
          doi = {10.1111/j.1365-2966.2012.21591.x},
archivePrefix = {arXiv},
       eprint = {1207.0682},
 primaryClass = {astro-ph.HE},
       adsurl = {https://ui.adsabs.harvard.edu/abs/2012MNRAS.425.2436C}
}

@ARTICLE{Grove1998ApJ...500..899G,
       author = {{Grove}, J.~E. and {Johnson}, W.~N. and {Kroeger}, R.~A. and {McNaron-Brown}, K. and {Skibo}, J.~G. and {Phlips}, B.~F.},
        title = "{Gamma-Ray Spectral States of Galactic Black Hole Candidates}",
      journal = {\apj},
         year = 1998,
        month = jun,
       volume = {500},
       number = {2},
        pages = {899-908},
          doi = {10.1086/305746},
archivePrefix = {arXiv},
       eprint = {astro-ph/9802242},
 primaryClass = {astro-ph},
       adsurl = {https://ui.adsabs.harvard.edu/abs/1998ApJ...500..899G}
}

@ARTICLE{Plotkin2013,
       author = {{Plotkin}, Richard. M. and {Gallo}, Elena and {Jonker}, Peter G.},
        title = "{The X-Ray Spectral Evolution of Galactic Black Hole X-Ray Binaries toward Quiescence}",
      journal = {\apj},
         year = 2013,
        month = aug,
       volume = {773},
       number = {1},
          eid = {59},
        pages = {59},
          doi = {10.1088/0004-637X/773/1/59},
archivePrefix = {arXiv},
       eprint = {1306.1570},
 primaryClass = {astro-ph.HE},
       adsurl = {https://ui.adsabs.harvard.edu/abs/2013ApJ...773...59P}
}

@ARTICLE{Haardt1993ApJ...413..507H,
       author = {{Haardt}, Francesco and {Maraschi}, Laura},
        title = "{X-Ray Spectra from Two-Phase Accretion Disks}",
      journal = {\apj},
         year = 1993,
        month = aug,
       volume = {413},
        pages = {507},
          doi = {10.1086/173020},
       adsurl = {https://ui.adsabs.harvard.edu/abs/1993ApJ...413..507H}
}

@ARTICLE{Shapiro1976ApJ...204..187S,
       author = {{Shapiro}, S.~L. and {Lightman}, A.~P. and {Eardley}, D.~M.},
        title = "{A two-temperature accretion disk model for Cygnus X-1: structure and spectrum.}",
      journal = {\apj},
         year = 1976,
        month = feb,
       volume = {204},
        pages = {187-199},
          doi = {10.1086/154162},
       adsurl = {https://ui.adsabs.harvard.edu/abs/1976ApJ...204..187S}
}

@ARTICLE{Sobolewska2011MNRAS.417..280S,
       author = {{Sobolewska}, M.~A. and {Papadakis}, I.~E. and {Done}, C. and {Malzac}, J.},
        title = "{Evidence for a change in the X-ray radiation mechanism in the hard state of Galactic black holes}",
      journal = {\mnras},
         year = 2011,
        month = oct,
       volume = {417},
       number = {1},
        pages = {280-288},
          doi = {10.1111/j.1365-2966.2011.19209.x},
archivePrefix = {arXiv},
       eprint = {1106.1645},
 primaryClass = {astro-ph.HE},
       adsurl = {https://ui.adsabs.harvard.edu/abs/2011MNRAS.417..280S}
}

@ARTICLE{Wilkins2011MNRAS.414.1269W,
       author = {{Wilkins}, D.~R. and {Fabian}, A.~C.},
        title = "{Determination of the X-ray reflection emissivity profile of 1H 0707-495}",
      journal = {\mnras},
         year = 2011,
        month = jun,
       volume = {414},
       number = {2},
        pages = {1269-1277},
          doi = {10.1111/j.1365-2966.2011.18458.x},
archivePrefix = {arXiv},
       eprint = {1102.0433},
 primaryClass = {astro-ph.HE},
       adsurl = {https://ui.adsabs.harvard.edu/abs/2011MNRAS.414.1269W}
}

@ARTICLE{Fabian2009Natur.459..540F,
       author = {{Fabian}, A.~C. and {Zoghbi}, A. and {Ross}, R.~R. and {Uttley}, P. and {Gallo}, L.~C. and {Brandt}, W.~N. and {Blustin}, A.~J. and {Boller}, T. and {Caballero-Garcia}, M.~D. and {Larsson}, J. and {Miller}, J.~M. and {Miniutti}, G. and {Ponti}, G. and {Reis}, R.~C. and {Reynolds}, C.~S. and {Tanaka}, Y. and {Young}, A.~J.},
        title = "{Broad line emission from iron K- and L-shell transitions in the active galaxy 1H0707-495}",
      journal = {\nat},
         year = 2009,
        month = may,
       volume = {459},
       number = {7246},
        pages = {540-542},
          doi = {10.1038/nature08007},
       adsurl = {https://ui.adsabs.harvard.edu/abs/2009Natur.459..540F}
}

@ARTICLE{Liu2023ApJ...950....5L,
       author = {{Liu}, Honghui and {Bambi}, Cosimo and {Jiang}, Jiachen and {Garc{\'\i}a}, Javier A. and {Ji}, Long and {Kong}, Lingda and {Ren}, Xiaoqin and {Zhang}, Shu and {Zhang}, Shuangnan},
        title = "{The Hard-to-soft Transition of GX 339-4 as Seen by Insight-HXMT}",
      journal = {\apj},
         year = 2023,
        month = jun,
       volume = {950},
       number = {1},
          eid = {5},
        pages = {5},
          doi = {10.3847/1538-4357/acca17},
archivePrefix = {arXiv},
       eprint = {2211.09543},
 primaryClass = {astro-ph.HE},
       adsurl = {https://ui.adsabs.harvard.edu/abs/2023ApJ...950....5L}
}

@ARTICLE{Reynolds2021,
       author = {{Reynolds}, Christopher S.},
        title = "{Observational Constraints on Black Hole Spin}",
      journal = {\araa},
         year = 2021,
        month = sep,
       volume = {59},
        pages = {117-154},
          doi = {10.1146/annurev-astro-112420-035022},
archivePrefix = {arXiv},
       eprint = {2011.08948},
 primaryClass = {astro-ph.HE},
       adsurl = {https://ui.adsabs.harvard.edu/abs/2021ARA&A..59..117R}
}

@ARTICLE{Walton2016,
       author = {{Walton}, D.~J. and {Tomsick}, J.~A. and {Madsen}, K.~K. and {Grinberg}, V. and {Barret}, D. and {Boggs}, S.~E. and {Christensen}, F.~E. and {Clavel}, M. and {Craig}, W.~W. and {Fabian}, A.~C. and {Fuerst}, F. and {Hailey}, C.~J. and {Harrison}, F.~A. and {Miller}, J.~M. and {Parker}, M.~L. and {Rahoui}, F. and {Stern}, D. and {Tao}, L. and {Wilms}, J. and {Zhang}, W.},
        title = "{The Soft State of Cygnus X-1 Observed with NuSTAR: A Variable Corona and a Stable Inner Disk}",
      journal = {\apj},
         year = 2016,
        month = jul,
       volume = {826},
       number = {1},
          eid = {87},
        pages = {87},
          doi = {10.3847/0004-637X/826/1/87},
archivePrefix = {arXiv},
       eprint = {1605.03966},
 primaryClass = {astro-ph.HE},
       adsurl = {https://ui.adsabs.harvard.edu/abs/2016ApJ...826...87W}
}

@ARTICLE{Wilkins2012MNRAS.424.1284W,
       author = {{Wilkins}, D.~R. and {Fabian}, A.~C.},
        title = "{Understanding X-ray reflection emissivity profiles in AGN: locating the X-ray source}",
      journal = {\mnras},
         year = 2012,
        month = aug,
       volume = {424},
       number = {2},
        pages = {1284-1296},
          doi = {10.1111/j.1365-2966.2012.21308.x},
archivePrefix = {arXiv},
       eprint = {1205.3179},
 primaryClass = {astro-ph.HE},
       adsurl = {https://ui.adsabs.harvard.edu/abs/2012MNRAS.424.1284W}
}

@ARTICLE{Greiner1996,
       author = {{Greiner}, J. and {Dennerl}, K. and {Predehl}, P.},
        title = "{ROSAT observation of GRS 1739-278.}",
      journal = {\aap},
         year = 1996,
        month = oct,
       volume = {314},
        pages = {L21-L24},
archivePrefix = {arXiv},
       eprint = {astro-ph/9608184},
 primaryClass = {astro-ph},
       adsurl = {https://ui.adsabs.harvard.edu/abs/1996A&A...314L..21G}
}

@ARTICLE{Chauhan2019,
       author = {{Chauhan}, J. and {Miller-Jones}, J.~C.~A. and {Anderson}, G.~E. and {Raja}, W. and {Bahramian}, A. and {Hotan}, A. and {Indermuehle}, B. and {Whiting}, M. and {Allison}, J.~R. and {Anderson}, C. and {Bunton}, J. and {Koribalski}, B. and {Mahony}, E.},
        title = "{An H I absorption distance to the black hole candidate X-ray binary MAXI J1535-571}",
      journal = {\mnras},
         year = 2019,
        month = sep,
       volume = {488},
       number = {1},
        pages = {L129-L133},
          doi = {10.1093/mnrasl/slz113},
archivePrefix = {arXiv},
       eprint = {1905.08497},
 primaryClass = {astro-ph.HE},
       adsurl = {https://ui.adsabs.harvard.edu/abs/2019MNRAS.488L.129C}
}

@ARTICLE{Xu2018maxij1535,
       author = {{Xu}, Yanjun and {Harrison}, Fiona A. and {Garc{\'\i}a}, Javier A. and {Fabian}, Andrew C. and {F{\"u}rst}, Felix and {Gandhi}, Poshak and {Grefenstette}, Brian W. and {Madsen}, Kristin K. and {Miller}, Jon M. and {Parker}, Michael L. and {Tomsick}, John A. and {Walton}, Dominic J.},
        title = "{Reflection Spectra of the Black Hole Binary Candidate MAXI J1535-571 in the Hard State Observed by NuSTAR}",
      journal = {\apjl},
         year = 2018,
        month = jan,
       volume = {852},
       number = {2},
          eid = {L34},
        pages = {L34},
          doi = {10.3847/2041-8213/aaa4b2},
archivePrefix = {arXiv},
       eprint = {1711.01346},
 primaryClass = {astro-ph.HE},
       adsurl = {https://ui.adsabs.harvard.edu/abs/2018ApJ...852L..34X}
}

@ARTICLE{Corral-Santana2016,
       author = {{Corral-Santana}, J.~M. and {Casares}, J. and {Mu{\~n}oz-Darias}, T. and {Bauer}, F.~E. and {Mart{\'\i}nez-Pais}, I.~G. and {Russell}, D.~M.},
        title = "{BlackCAT: A catalogue of stellar-mass black holes in X-ray transients}",
      journal = {\aap},
         year = 2016,
        month = mar,
       volume = {587},
          eid = {A61},
        pages = {A61},
          doi = {10.1051/0004-6361/201527130},
archivePrefix = {arXiv},
       eprint = {1510.08869},
 primaryClass = {astro-ph.HE},
       adsurl = {https://ui.adsabs.harvard.edu/abs/2016A&A...587A..61C}
}

@ARTICLE{Iyer2015,
       author = {{Iyer}, N. and {Nandi}, A. and {Mandal}, S.},
        title = "{Determination of the Mass of IGR J17091-3624 from ``Spectro-temporal'' Variations during the Onset Phase of the 2011 Outburst}",
      journal = {\apj},
         year = 2015,
        month = jul,
       volume = {807},
       number = {1},
          eid = {108},
        pages = {108},
          doi = {10.1088/0004-637X/807/1/108},
archivePrefix = {arXiv},
       eprint = {1505.02529},
 primaryClass = {astro-ph.HE},
       adsurl = {https://ui.adsabs.harvard.edu/abs/2015ApJ...807..108I}
}

@ARTICLE{Bardeen1975ApJ...195L..65B,
       author = {{Bardeen}, James M. and {Petterson}, Jacobus A.},
        title = "{The Lense-Thirring Effect and Accretion Disks around Kerr Black Holes}",
      journal = {\apjl},
         year = 1975,
        month = jan,
       volume = {195},
        pages = {L65},
          doi = {10.1086/181711},
       adsurl = {https://ui.adsabs.harvard.edu/abs/1975ApJ...195L..65B}
}

@ARTICLE{ixpe,
       author = {{Weisskopf}, Martin C. and {Soffitta}, Paolo and {Baldini}, Luca and {Ramsey}, Brian D. and {O'Dell}, Stephen L. and {Romani}, Roger W. and {Matt}, Giorgio and {Deininger}, William D. and {Baumgartner}, Wayne H. and {Bellazzini}, Ronaldo and {Costa}, Enrico and {Kolodziejczak}, Jeffery J. and {Latronico}, Luca and {Marshall}, Herman L. and {Muleri}, Fabio and {Bongiorno}, Stephen D. and {Tennant}, Allyn and {Bucciantini}, Niccolo and {Dovciak}, Michal and {Marin}, Frederic and {Marscher}, Alan and {Poutanen}, Juri and {Slane}, Pat and {Turolla}, Roberto and {Kalinowski}, William and {Di Marco}, Alessandro and {Fabiani}, Sergio and {Minuti}, Massimo and {La Monaca}, Fabio and {Pinchera}, Michele and {Rankin}, John and {Sgro'}, Carmelo and {Trois}, Alessio and {Xie}, Fei and {Alexander}, Cheryl and {Allen}, D. Zachery and {Amici}, Fabrizio and {Andersen}, Jason and {Antonelli}, Angelo and {Antoniak}, Spencer and {Attina'}, Primo and {Barbanera}, Mattia and {Bachetti}, Matteo and {Baggett}, Randy M. and {Bladt}, Jeff and {Brez}, Alessandro and {Bonino}, Raffaella and {Boree}, Christopher and {Borotto}, Fabio and {Breeding}, Shawn and {Brienza}, Daniele and {Bygott}, H. Kyle and {Caporale}, Ciro and {Cardelli}, Claudia and {Carpentiero}, Rita and {Castellano}, Simone and {Castronuovo}, Marco and {Cavalli}, Luca and {Cavazzuti}, Elisabetta and {Ceccanti}, Marco and {Centrone}, Mauro and {Citraro}, Saverio and {D'Amico}, Fabio and {D'Alba}, Elisa and {Di Gesu}, Laura and {Del Monte}, Ettore and {Dietz}, Kurtis L. and {Di Lalla}, Niccolo' and {Di Persio}, Giuseppe and {Dolan}, David and {Donnarumma}, Immacolata and {Evangelista}, Yuri and {Ferrant}, Kevin and {Ferrazzoli}, Riccardo and {Ferrie}, MacKenzie and {Footdale}, Joseph and {Forsyth}, Brent and {Foster}, Michelle and {Garelick}, Benjamin and {Gunji}, Shuichi and {Gurnee}, Eli and {Head}, Michael and {Hibbard}, Grant and {Johnson}, Samantha and {Kelly}, Erik and {Kilaru}, Kiranmayee and {Lefevre}, Carlo and {Le Roy}, Shelley and {Loffredo}, Pasqualino and {Lorenzi}, Paolo and {Lucchesi}, Leonardo and {Maddox}, Tyler and {Magazzu}, Guido and {Maldera}, Simone and {Manfreda}, Alberto and {Mangraviti}, Elio and {Marengo}, Marco and {Marrocchesi}, Alessandra and {Massaro}, Francesco and {Mauger}, David and {McCracken}, Jeffrey and {McEachen}, Michael and {Mize}, Rondal and {Mereu}, Paolo and {Mitchell}, Scott and {Mitsuishi}, Ikuyuki and {Morbidini}, Alfredo and {Mosti}, Federico and {Nasimi}, Hikmat and {Negri}, Barbara and {Negro}, Michela and {Nguyen}, Toan and {Nitschke}, Isaac and {Nuti}, Alessio and {Onizuka}, Mitch and {Oppedisano}, Chiara and {Orsini}, Leonardo and {Osborne}, Darren and {Pacheco}, Richard and {Paggi}, Alessandro and {Painter}, Will and {Pavelitz}, Steven D. and {Pentz}, Christina and {Piazzolla}, Raffaele and {Perri}, Matteo and {Pesce-Rollins}, Melissa and {Peterson}, Colin and {Pilia}, Maura and {Profeti}, Alessandro and {Puccetti}, Simonetta and {Ranganathan}, Jaganathan and {Ratheesh}, Ajay and {Reedy}, Lee and {Root}, Noah and {Rubini}, Alda and {Ruswick}, Stephanie and {Sanchez}, Javier and {Sarra}, Paolo and {Santoli}, Francesco and {Scalise}, Emanuele and {Sciortino}, Andrea and {Schroeder}, Christopher and {Seek}, Tim and {Sosdian}, Kalie and {Spandre}, Gloria and {Speegle}, Chet O. and {Tamagawa}, Toru and {Tardiola}, Marcello and {Tobia}, Antonino and {Thomas}, Nicholas E. and {Valerie}, Robert and {Vimercati}, Marco and {Walden}, Amy L. and {Weddendorf}, Bruce and {Wedmore}, Jeffrey and {Welch}, David and {Zanetti}, Davide and {Zanetti}, Francesco},
        title = "{The Imaging X-Ray Polarimetry Explorer (IXPE): Pre-Launch}",
      journal = {Journal of Astronomical Telescopes, Instruments, and Systems},
         year = 2022,
        month = apr,
       volume = {8},
       number = {2},
          eid = {026002},
        pages = {026002},
          doi = {10.1117/1.JATIS.8.2.026002},
       adsurl = {https://ui.adsabs.harvard.edu/abs/2022JATIS...8b6002W}
}

@INPROCEEDINGS{nicer,
       author = {{Gendreau}, Keith C. and {Arzoumanian}, Zaven and {Okajima}, Takashi},
        title = "{The Neutron star Interior Composition ExploreR (NICER): an Explorer mission of opportunity for soft x-ray timing spectroscopy}",
    booktitle = {Space Telescopes and Instrumentation 2012: Ultraviolet to Gamma Ray},
         year = 2012,
       editor = {{Takahashi}, Tadayuki and {Murray}, Stephen S. and {den Herder}, Jan-Willem A.},
       series = {\procspie},
       volume = {8443},
        month = sep,
          eid = {844313},
        pages = {844313},
          doi = {10.1117/12.926396},
       adsurl = {https://ui.adsabs.harvard.edu/abs/2012SPIE.8443E..13G}
}

@ARTICLE{nustar,
   author = {{Harrison}, F.~A. and {Craig}, W.~W. and {Christensen}, F.~E. and 
	{Hailey}, C.~J. and {Zhang}, W.~W. and {Boggs}, S.~E. and {Stern}, D. and 
	{Cook}, W.~R. and {Forster}, K. and {Giommi}, P. and {Grefenstette}, B.~W. and 
	{Kim}, Y. and {Kitaguchi}, T. and {Koglin}, J.~E. and {Madsen}, K.~K. and 
	{Mao}, P.~H. and {Miyasaka}, H. and {Mori}, K. and {Perri}, M. and 
	{Pivovaroff}, M.~J. and {Puccetti}, S. and {Rana}, V.~R. and 
	{Westergaard}, N.~J. and {Willis}, J. and {Zoglauer}, A. and 
	{An}, H. and {Bachetti}, M. and {Barri{\`e}re}, N.~M. and {Bellm}, E.~C. and 
	{Bhalerao}, V. and {Brejnholt}, N.~F. and {Fuerst}, F. and {Liebe}, C.~C. and 
	{Markwardt}, C.~B. and {Nynka}, M. and {Vogel}, J.~K. and {Walton}, D.~J. and 
	{Wik}, D.~R. and {Alexander}, D.~M. and {Cominsky}, L.~R. and 
	{Hornschemeier}, A.~E. and {Hornstrup}, A. and {Kaspi}, V.~M. and 
	{Madejski}, G.~M. and {Matt}, G. and {Molendi}, S. and {Smith}, D.~M. and 
	{Tomsick}, J.~A. and {Ajello}, M. and {Ballantyne}, D.~R. and 
	{Balokovi{\'c}}, M. and {Barret}, D. and {Bauer}, F.~E. and 
	{Blandford}, R.~D. and {Niel Brandt}, W. and {Brenneman}, L.~W. and 
	{Chiang}, J. and {Chakrabarty}, D. and {Chenevez}, J. and {Comastri}, A. and 
	{Dufour}, F. and {Elvis}, M. and {Fabian}, A.~C. and {Farrah}, D. and 
	{Fryer}, C.~L. and {Gotthelf}, E.~V. and {Grindlay}, J.~E. and 
	{Helfand}, D.~J. and {Krivonos}, R. and {Meier}, D.~L. and {Miller}, J.~M. and 
	{Natalucci}, L. and {Ogle}, P. and {Ofek}, E.~O. and {Ptak}, A. and 
	{Reynolds}, S.~P. and {Rigby}, J.~R. and {Tagliaferri}, G. and 
	{Thorsett}, S.~E. and {Treister}, E. and {Urry}, C.~M.},
    title = "{The Nuclear Spectroscopic Telescope Array (NuSTAR) High-energy X-Ray Mission}",
  journal = {\apj},
archivePrefix = "arXiv",
   eprint = {1301.7307},
 primaryClass = "astro-ph.IM",
     year = 2013,
    month = jun,
   volume = 770,
      eid = {103},
    pages = {103},
      doi = {10.1088/0004-637X/770/2/103},
   adsurl = {http://adsabs.harvard.edu/abs/2013ApJ...770..103H}
}

@INCOLLECTION{McClintock2006csxs.book..157M,
       author = {{McClintock}, Jeffrey E. and {Remillard}, Ronald A.},
        title = "{Black hole binaries}",
    booktitle = {Compact stellar X-ray sources}, 
    series={Cambridge Astrophysics Series},  
    publisher = {Cambridge University Press},
 address = {Cambridge}, 
         year = 2006,
    editor = {{Lewin}, W. and {van der Klis}, M.},
       volume = {39},
        pages = {157-213},
          doi = {10.48550/arXiv.astro-ph/0306213},
       adsurl = {https://ui.adsabs.harvard.edu/abs/2006csxs.book..157M}
}

@ARTICLE{Shakura1973,
       author = {{Shakura}, N.~I. and {Sunyaev}, R.~A.},
        title = "{Black holes in binary systems. Observational appearance.}",
      journal = {\aap},
         year = 1973,
        month = jan,
       volume = {24},
        pages = {337-355},
       adsurl = {https://ui.adsabs.harvard.edu/abs/1973A&A....24..337S}
}

@ARTICLE{Matt1993MNRAS.260..663M,
       author = {{Matt}, Giorgio},
        title = "{X-ray polarization properties of a centrally illuminated accretion disc.}",
      journal = {\mnras},
         year = 1993,
        month = feb,
       volume = {260},
        pages = {663-674},
          doi = {10.1093/mnras/260.3.663},
       adsurl = {https://ui.adsabs.harvard.edu/abs/1993MNRAS.260..663M}
}

@ARTICLE{Marra2024A&A...684A..95M,
       author = {{Marra}, L. and {Brigitte}, M. and {Rodriguez Cavero}, N. and {Chun}, S. and {Steiner}, J.~F. and {Dov{\v{c}}iak}, M. and {Nowak}, M. and {Bianchi}, S. and {Capitanio}, F. and {Ingram}, A. and {Matt}, G. and {Muleri}, F. and {Podgorn{\'y}}, J. and {Poutanen}, J. and {Svoboda}, J. and {Taverna}, R. and {Ursini}, F. and {Veledina}, A. and {De Rosa}, A. and {Garc{\'\i}a}, J.~A. and {Lutovinov}, A.~A. and {Mereminskiy}, I.~A. and {Farinelli}, R. and {Gunji}, S. and {Kaaret}, P. and {Kallman}, T. and {Krawczynski}, H. and {Kan}, Y. and {Hu}, K. and {Marinucci}, A. and {Mastroserio}, G. and {Mikus̆incov{\'a}}, R. and {Parra}, M. and {Petrucci}, P. -O. and {Ratheesh}, A. and {Soffitta}, P. and {Tombesi}, F. and {Zane}, S. and {Agudo}, I. and {Antonelli}, L.~A. and {Bachetti}, M. and {Baldini}, L. and {Baumgartner}, W.~H. and {Bellazzini}, R. and {Bongiorno}, S.~D. and {Bonino}, R. and {Brez}, A. and {Bucciantini}, N. and {Castellano}, S. and {Cavazzuti}, E. and {Chen}, C. and {Ciprini}, S. and {Costa}, E. and {Del Monte}, E. and {Di Gesu}, L. and {Di Lalla}, N. and {Di Marco}, A. and {Donnarumma}, I. and {Doroshenko}, V. and {Ehlert}, S.~R. and {Enoto}, T. and {Evangelista}, Y. and {Fabiani}, S. and {Ferrazzoli}, R. and {Hayashida}, K. and {Heyl}, J. and {Iwakiri}, W. and {Jorstad}, S.~G. and {Karas}, V. and {Kislat}, F. and {Kitaguchi}, T. and {Kolodziejczak}, J.~J. and {La Monaca}, F. and {Latronico}, L. and {Liodakis}, I. and {Maldera}, S. and {Manfreda}, A. and {Marin}, F. and {Marscher}, A.~P. and {Marshall}, H.~L. and {Massaro}, F. and {Mitsuishi}, I. and {Mizuno}, T. and {Negro}, M. and {Ng}, C.~Y. and {O'Dell}, S.~L. and {Omodei}, N. and {Oppedisano}, C. and {Papitto}, A. and {Pavlov}, G.~G. and {Peirson}, A.~L. and {Perri}, M. and {Pesce-Rollins}, M. and {Pilia}, M. and {Possenti}, A. and {Puccetti}, S. and {Ramsey}, B.~D. and {Rankin}, J. and {Roberts}, O.~J. and {Romani}, R.~W. and {Sgr{\`o}}, C. and {Slane}, P. and {Spandre}, G. and {Swartz}, D.~A. and {Tamagawa}, T. and {Tavecchio}, F. and {Tawara}, Y. and {Tennant}, A.~F. and {Thomas}, N.~E. and {Trois}, A. and {Tsygankov}, S.~S. and {Turolla}, R. and {Vink}, J. and {Weisskopf}, M.~C. and {Wu}, K. and {Xie}, F.},
        title = "{IXPE observation confirms a high spin in the accreting black hole 4U 1957+115}",
      journal = {\aap},
         year = 2024,
        month = apr,
       volume = {684},
          eid = {A95},
        pages = {A95},
          doi = {10.1051/0004-6361/202348277},
archivePrefix = {arXiv},
       eprint = {2310.11125},
 primaryClass = {astro-ph.HE},
       adsurl = {https://ui.adsabs.harvard.edu/abs/2024A&A...684A..95M}
}

@ARTICLE{Podgorny2023MNRAS.524.3853P,
       author = {{Podgorn{\'y}}, J. and {Dov{\v{c}}iak}, M. and {Goosmann}, R. and {Marin}, F. and {Matt}, G. and {R{\'o}{\.z}a{\'n}ska}, A. and {Karas}, V.},
        title = "{Spectral and polarization properties of reflected X-ray emission from black-hole accretion discs for a distant observer: the lamp-post model}",
      journal = {\mnras},
         year = 2023,
        month = sep,
       volume = {524},
       number = {3},
        pages = {3853-3876},
          doi = {10.1093/mnras/stad2169},
archivePrefix = {arXiv},
       eprint = {2307.08819},
 primaryClass = {astro-ph.HE},
       adsurl = {https://ui.adsabs.harvard.edu/abs/2023MNRAS.524.3853P}
}

@ARTICLE{Podgorny2022MNRAS.510.4723P,
       author = {{Podgorn{\'y}}, J. and {Dov{\v{c}}iak}, M. and {Marin}, F. and {Goosmann}, R. and {R{\'o}{\.z}a{\'n}ska}, A.},
        title = "{Spectral and polarization properties of reflected X-ray emission from black hole accretion discs}",
      journal = {\mnras},
         year = 2022,
        month = mar,
       volume = {510},
       number = {4},
        pages = {4723-4735},
          doi = {10.1093/mnras/stab3714},
archivePrefix = {arXiv},
       eprint = {2201.07494},
 primaryClass = {astro-ph.HE},
       adsurl = {https://ui.adsabs.harvard.edu/abs/2022MNRAS.510.4723P}
}

@ARTICLE{Dovvciak2008MNRAS.391...32D,
       author = {{Dov{\v{c}}iak}, M. and {Muleri}, F. and {Goosmann}, R.~W. and {Karas}, V. and {Matt}, G.},
        title = "{Thermal disc emission from a rotating black hole: X-ray polarization signatures}",
      journal = {\mnras},
         year = 2008,
        month = nov,
       volume = {391},
       number = {1},
        pages = {32-38},
          doi = {10.1111/j.1365-2966.2008.13872.x},
archivePrefix = {arXiv},
       eprint = {0809.0418},
 primaryClass = {astro-ph},
       adsurl = {https://ui.adsabs.harvard.edu/abs/2008MNRAS.391...32D}
}

@ARTICLE{Weisskopf2022JATIS...8b6002W,
       author = {{Weisskopf}, Martin C. and {Soffitta}, Paolo and {Baldini}, Luca and {Ramsey}, Brian D. and {O'Dell}, Stephen L. and {Romani}, Roger W. and {Matt}, Giorgio and {Deininger}, William D. and {Baumgartner}, Wayne H. and {Bellazzini}, Ronaldo and {Costa}, Enrico and {Kolodziejczak}, Jeffery J. and {Latronico}, Luca and {Marshall}, Herman L. and {Muleri}, Fabio and {Bongiorno}, Stephen D. and {Tennant}, Allyn and {Bucciantini}, Niccolo and {Dovciak}, Michal and {Marin}, Frederic and {Marscher}, Alan and {Poutanen}, Juri and {Slane}, Pat and {Turolla}, Roberto and {Kalinowski}, William and {Di Marco}, Alessandro and {Fabiani}, Sergio and {Minuti}, Massimo and {La Monaca}, Fabio and {Pinchera}, Michele and {Rankin}, John and {Sgro'}, Carmelo and {Trois}, Alessio and {Xie}, Fei and {Alexander}, Cheryl and {Allen}, D. Zachery and {Amici}, Fabrizio and {Andersen}, Jason and {Antonelli}, Angelo and {Antoniak}, Spencer and {Attin{\`a}}, Primo and {Barbanera}, Mattia and {Bachetti}, Matteo and {Baggett}, Randy M. and {Bladt}, Jeff and {Brez}, Alessandro and {Bonino}, Raffaella and {Boree}, Christopher and {Borotto}, Fabio and {Breeding}, Shawn and {Brienza}, Daniele and {Bygott}, H. Kyle and {Caporale}, Ciro and {Cardelli}, Claudia and {Carpentiero}, Rita and {Castellano}, Simone and {Castronuovo}, Marco and {Cavalli}, Luca and {Cavazzuti}, Elisabetta and {Ceccanti}, Marco and {Centrone}, Mauro and {Citraro}, Saverio and {D'Amico}, Fabio and {D'Alba}, Elisa and {Di Gesu}, Laura and {Del Monte}, Ettore and {Dietz}, Kurtis L. and {Di Lalla}, Niccolo' and {Persio}, Giuseppe Di and {Dolan}, David and {Donnarumma}, Immacolata and {Evangelista}, Yuri and {Ferrant}, Kevin and {Ferrazzoli}, Riccardo and {Ferrie}, MacKenzie and {Footdale}, Joseph and {Forsyth}, Brent and {Foster}, Michelle and {Garelick}, Benjamin and {Gunji}, Shuichi and {Gurnee}, Eli and {Head}, Michael and {Hibbard}, Grant and {Johnson}, Samantha and {Kelly}, Erik and {Kilaru}, Kiranmayee and {Lefevre}, Carlo and {Roy}, Shelley Le and {Loffredo}, Pasqualino and {Lorenzi}, Paolo and {Lucchesi}, Leonardo and {Maddox}, Tyler and {Magazzu}, Guido and {Maldera}, Simone and {Manfreda}, Alberto and {Mangraviti}, Elio and {Marengo}, Marco and {Marrocchesi}, Alessandra and {Massaro}, Francesco and {Mauger}, David and {McCracken}, Jeffrey and {McEachen}, Michael and {Mize}, Rondal and {Mereu}, Paolo and {Mitchell}, Scott and {Mitsuishi}, Ikuyuki and {Morbidini}, Alfredo and {Mosti}, Federico and {Nasimi}, Hikmat and {Negri}, Barbara and {Negro}, Michela and {Nguyen}, Toan and {Nitschke}, Isaac and {Nuti}, Alessio and {Onizuka}, Mitch and {Oppedisano}, Chiara and {Orsini}, Leonardo and {Osborne}, Darren and {Pacheco}, Richard and {Paggi}, Alessandro and {Painter}, Will and {Pavelitz}, Steven D. and {Pentz}, Christina and {Piazzolla}, Raffaele and {Perri}, Matteo and {Pesce-Rollins}, Melissa and {Peterson}, Colin and {Pilia}, Maura and {Profeti}, Alessandro and {Puccetti}, Simonetta and {Ranganathan}, Jaganathan and {Ratheesh}, Ajay and {Reedy}, Lee and {Root}, Noah and {Rubini}, Alda and {Ruswick}, Stephanie and {Sanchez}, Javier and {Sarra}, Paolo and {Santoli}, Francesco and {Scalise}, Emanuele and {Sciortino}, Andrea and {Schroeder}, Christopher and {Seek}, Tim and {Sosdian}, Kalie and {Spandre}, Gloria and {Speegle}, Chet O. and {Tamagawa}, Toru and {Tardiola}, Marcello and {Tobia}, Antonino and {Thomas}, Nicholas E. and {Valerie}, Robert and {Vimercati}, Marco and {Walden}, Amy L. and {Weddendorf}, Bruce and {Wedmore}, Jeffrey and {Welch}, David and {Zanetti}, Davide and {Zanetti}, Francesco},
        title = "{The Imaging X-Ray Polarimetry Explorer (IXPE): Pre-Launch}",
      journal = {Journal of Astronomical Telescopes, Instruments, and Systems},
         year = 2022,
        month = apr,
       volume = {8},
       number = {2},
          eid = {026002},
        pages = {026002},
          doi = {10.1117/1.JATIS.8.2.026002},
archivePrefix = {arXiv},
       eprint = {2112.01269},
 primaryClass = {astro-ph.IM},
       adsurl = {https://ui.adsabs.harvard.edu/abs/2022JATIS...8b6002W}
}

@ARTICLE{Garcia2022ApJ...926...13G,
       author = {{Garc{\'\i}a}, Javier A. and {Dauser}, Thomas and {Ludlam}, Renee and {Parker}, Michael and {Fabian}, Andrew and {Harrison}, Fiona A. and {Wilms}, J{\"o}rn},
        title = "{Relativistic X-Ray Reflection Models for Accreting Neutron Stars}",
      journal = {\apj},
         year = 2022,
        month = feb,
       volume = {926},
       number = {1},
          eid = {13},
        pages = {13},
          doi = {10.3847/1538-4357/ac3cb7},
archivePrefix = {arXiv},
       eprint = {2111.12838},
 primaryClass = {astro-ph.HE},
       adsurl = {https://ui.adsabs.harvard.edu/abs/2022ApJ...926...13G}
}

@ARTICLE{Ingram2024ApJ...968...76I,
       author = {{Ingram}, Adam and {Bollemeijer}, Niek and {Veledina}, Alexandra and {Dov{\v{c}}iak}, Michal and {Poutanen}, Juri and {Egron}, Elise and {Russell}, Thomas D. and {Trushkin}, Sergei A. and {Negro}, Michela and {Ratheesh}, Ajay and {Capitanio}, Fiamma and {Connors}, Riley and {Neilsen}, Joseph and {Kraus}, Alexander and {Iacolina}, Maria Noemi and {Pellizzoni}, Alberto and {Pilia}, Maura and {Carotenuto}, Francesco and {Matt}, Giorgio and {Mastroserio}, Guglielmo and {Kaaret}, Philip and {Bianchi}, Stefano and {Garc{\'\i}a}, Javier A. and {Bachetti}, Matteo and {Wu}, Kinwah and {Costa}, Enrico and {Ewing}, Melissa and {Kravtsov}, Vadim and {Krawczynski}, Henric and {Loktev}, Vladislav and {Marinucci}, Andrea and {Marra}, Lorenzo and {Miku{\v{s}}incov{\'a}}, Romana and {Nathan}, Edward and {Parra}, Maxime and {Petrucci}, Pierre-Olivier and {Righini}, Simona and {Soffitta}, Paolo and {Steiner}, James F. and {Svoboda}, Ji{\v{r}}{\'\i} and {Tombesi}, Francesco and {Tugliani}, Stefano and {Ursini}, Francesco and {Yang}, Yi-Jung and {Zane}, Silvia and {Zhang}, Wenda and {Agudo}, Iv{\'a}n and {Antonelli}, Lucio A. and {Baldini}, Luca and {Baumgartner}, Wayne H. and {Bellazzini}, Ronaldo and {Bongiorno}, Stephen D. and {Bonino}, Raffaella and {Brez}, Alessandro and {Bucciantini}, Niccol{\`o} and {Castellano}, Simone and {Cavazzuti}, Elisabetta and {Chen}, Chien-Ting and {Ciprini}, Stefano and {De Rosa}, Alessandra and {Del Monte}, Ettore and {Di Gesu}, Laura and {Di Lalla}, Niccol{\`o} and {Di Marco}, Alessandro and {Donnarumma}, Immacolata and {Doroshenko}, Victor and {Ehlert}, Steven R. and {Enoto}, Teruaki and {Evangelista}, Yuri and {Fabiani}, Sergio and {Ferrazzoli}, Riccardo and {Gunji}, Shuichi and {Hayashida}, Kiyoshi and {Heyl}, Jeremy and {Iwakiri}, Wataru and {Jorstad}, Svetlana G. and {Karas}, Vladimir and {Kislat}, Fabian and {Kitaguchi}, Takao and {Kolodziejczak}, Jeffery J. and {La Monaca}, Fabio and {Latronico}, Luca and {Liodakis}, Ioannis and {Maldera}, Simone and {Manfreda}, Alberto and {Marin}, Fr{\'e}d{\'e}ric and {Marscher}, Alan P. and {Marshall}, Herman L. and {Massaro}, Francesco and {Mitsuishi}, Ikuyuki and {Mizuno}, Tsunefumi and {Muleri}, Fabio and {Ng}, Chi-Yung and {O'Dell}, Stephen L. and {Omodei}, Nicola and {Oppedisano}, Chiara and {Papitto}, Alessandro and {Pavlov}, George G. and {Peirson}, Abel L. and {Perri}, Matteo and {Pesce-Rollins}, Melissa and {Possenti}, Andrea and {Puccetti}, Simonetta and {Ramsey}, Brian D. and {Rankin}, John and {Roberts}, Oliver J. and {Romani}, Roger W. and {Sgr{\`o}}, Carmelo and {Slane}, Patrick and {Spandre}, Gloria and {Swartz}, Douglas A. and {Tamagawa}, Toru and {Tavecchio}, Fabrizio and {Taverna}, Roberto and {Tawara}, Yuzuru and {Tennant}, Allyn F. and {Thomas}, Nicholas E. and {Trois}, Alessio and {Tsygankov}, Sergey S. and {Turolla}, Roberto and {Vink}, Jacco and {Weisskopf}, Martin C. and {Xie}, Fei and {IXPE Collaboration}},
        title = "{Tracking the X-Ray Polarization of the Black Hole Transient Swift J1727.8{\textendash}1613 during a State Transition}",
      journal = {\apj},
         year = 2024,
        month = jun,
       volume = {968},
       number = {2},
          eid = {76},
        pages = {76},
          doi = {10.3847/1538-4357/ad3faf},
archivePrefix = {arXiv},
       eprint = {2311.05497},
 primaryClass = {astro-ph.HE},
       adsurl = {https://ui.adsabs.harvard.edu/abs/2024ApJ...968...76I}
}

@ARTICLE{Svoboda2024ApJ...966L..35S,
       author = {{Svoboda}, Ji{\v{r}}{\'\i} and {Dov{\v{c}}iak}, Michal and {Steiner}, James F. and {Kaaret}, Philip and {Podgorn{\'y}}, Jakub and {Poutanen}, Juri and {Veledina}, Alexandra and {Muleri}, Fabio and {Taverna}, Roberto and {Krawczynski}, Henric and {Brigitte}, Ma{\"\i}mouna and {Datta}, Sudeb Ranjan and {Bianchi}, Stefano and {Mu{\~n}oz-Darias}, Teo and {Negro}, Michela and {Rodriguez Cavero}, Nicole and {Castro Segura}, Noel and {Bollemeijer}, Niek and {Garc{\'\i}a}, Javier A. and {Ingram}, Adam and {Matt}, Giorgio and {Nathan}, Edward and {Weisskopf}, Martin C. and {Altamirano}, Diego and {Baldini}, Luca and {Capitanio}, Fiamma and {Egron}, Elise and {Emami}, Razieh and {Hu}, Kun and {Marra}, Lorenzo and {Mastroserio}, Guglielmo and {Petrucci}, Pierre-Olivier and {Ratheesh}, Ajay and {Soffitta}, Paolo and {Tombesi}, Francesco and {Yang}, Yi-Jung and {Zhang}, Yuexin},
        title = "{Dramatic Drop in the X-Ray Polarization of Swift J1727.8{\textendash}1613 in the Soft Spectral State}",
      journal = {\apjl},
         year = 2024,
        month = may,
       volume = {966},
       number = {2},
          eid = {L35},
        pages = {L35},
          doi = {10.3847/2041-8213/ad402e},
archivePrefix = {arXiv},
       eprint = {2403.04689},
 primaryClass = {astro-ph.HE},
       adsurl = {https://ui.adsabs.harvard.edu/abs/2024ApJ...966L..35S}
}

@ARTICLE{Veledina2023ApJ...958L..16V,
       author = {{Veledina}, Alexandra and {Muleri}, Fabio and {Dov{\v{c}}iak}, Michal and {Poutanen}, Juri and {Ratheesh}, Ajay and {Capitanio}, Fiamma and {Matt}, Giorgio and {Soffitta}, Paolo and {Tennant}, Allyn F. and {Negro}, Michela and {Kaaret}, Philip and {Costa}, Enrico and {Ingram}, Adam and {Svoboda}, Ji{\v{r}}{\'\i} and {Krawczynski}, Henric and {Bianchi}, Stefano and {Steiner}, James F. and {Garc{\'\i}a}, Javier A. and {Kravtsov}, Vadim and {Nitindala}, Anagha P. and {Ewing}, Melissa and {Mastroserio}, Guglielmo and {Marinucci}, Andrea and {Ursini}, Francesco and {Tombesi}, Francesco and {Tsygankov}, Sergey S. and {Yang}, Yi-Jung and {Weisskopf}, Martin C. and {Trushkin}, Sergei A. and {Egron}, Elise and {Iacolina}, Maria Noemi and {Pilia}, Maura and {Marra}, Lorenzo and {Miku{\v{s}}incov{\'a}}, Romana and {Nathan}, Edward and {Parra}, Maxime and {Petrucci}, Pierre-Olivier and {Podgorn{\'y}}, Jakub and {Tugliani}, Stefano and {Zane}, Silvia and {Zhang}, Wenda and {Agudo}, Iv{\'a}n and {Antonelli}, Lucio A. and {Bachetti}, Matteo and {Baldini}, Luca and {Baumgartner}, Wayne H. and {Bellazzini}, Ronaldo and {Bongiorno}, Stephen D. and {Bonino}, Raffaella and {Brez}, Alessandro and {Bucciantini}, Niccol{\`o} and {Castellano}, Simone and {Cavazzuti}, Elisabetta and {Chen}, Chien-Ting and {Ciprini}, Stefano and {De Rosa}, Alessandra and {Del Monte}, Ettore and {Di Gesu}, Laura and {Di Lalla}, Niccol{\`o} and {Di Marco}, Alessandro and {Donnarumma}, Immacolata and {Doroshenko}, Victor and {Ehlert}, Steven R. and {Enoto}, Teruaki and {Evangelista}, Yuri and {Fabiani}, Sergio and {Ferrazzoli}, Riccardo and {Gunji}, Shuichi and {Hayashida}, Kiyoshi and {Heyl}, Jeremy and {Iwakiri}, Wataru and {Jorstad}, Svetlana G. and {Karas}, Vladimir and {Kislat}, Fabian and {Kitaguchi}, Takao and {Kolodziejczak}, Jeffery J. and {La Monaca}, Fabio and {Latronico}, Luca and {Liodakis}, Ioannis and {Maldera}, Simone and {Manfreda}, Alberto and {Marin}, Fr{\'e}d{\'e}ric and {Marscher}, Alan P. and {Marshall}, Herman L. and {Massaro}, Francesco and {Mitsuishi}, Ikuyuki and {Mizuno}, Tsunefumi and {Ng}, Chi-Yung and {O'Dell}, Stephen L. and {Omodei}, Nicola and {Oppedisano}, Chiara and {Papitto}, Alessandro and {Pavlov}, George G. and {Peirson}, Abel L. and {Perri}, Matteo and {Pesce-Rollins}, Melissa and {Possenti}, Andrea and {Puccetti}, Simonetta and {Ramsey}, Brian D. and {Rankin}, John and {Roberts}, Oliver J. and {Romani}, Roger W. and {Sgr{\`o}}, Carmelo and {Slane}, Patrick and {Spandre}, Gloria and {Swartz}, Douglas A. and {Tamagawa}, Toru and {Tavecchio}, Fabrizio and {Taverna}, Roberto and {Tawara}, Yuzuru and {Thomas}, Nicholas E. and {Trois}, Alessio and {Turolla}, Roberto and {Vink}, Jacco and {Wu}, Kinwah and {Xie}, Fei},
        title = "{Discovery of X-Ray Polarization from the Black Hole Transient Swift J1727.8-1613}",
      journal = {\apjl},
         year = 2023,
        month = nov,
       volume = {958},
       number = {1},
          eid = {L16},
        pages = {L16},
          doi = {10.3847/2041-8213/ad0781},
archivePrefix = {arXiv},
       eprint = {2309.15928},
 primaryClass = {astro-ph.HE},
       adsurl = {https://ui.adsabs.harvard.edu/abs/2023ApJ...958L..16V}
}

@ARTICLE{Podgorny2024A&A...686L..12P,
       author = {{Podgorn{\'y}}, J. and {Svoboda}, J. and {Dov{\v{c}}iak}, M. and {Veledina}, A. and {Poutanen}, J. and {Kaaret}, P. and {Bianchi}, S. and {Ingram}, A. and {Capitanio}, F. and {Datta}, S.~R. and {Egron}, E. and {Krawczynski}, H. and {Matt}, G. and {Muleri}, F. and {Petrucci}, P. -O. and {Russell}, T.~D. and {Steiner}, J.~F. and {Bollemeijer}, N. and {Brigitte}, M. and {Castro Segura}, N. and {Emami}, R. and {Garc{\'\i}a}, J.~A. and {Hu}, K. and {Iacolina}, M.~N. and {Kravtsov}, V. and {Marra}, L. and {Mastroserio}, G. and {Mu{\~n}oz-Darias}, T. and {Nathan}, E. and {Negro}, M. and {Ratheesh}, A. and {Rodriguez Cavero}, N. and {Taverna}, R. and {Tombesi}, F. and {Yang}, Y.~J. and {Zhang}, W. and {Zhang}, Y.},
        title = "{Recovery of the X-ray polarisation of Swift J1727.8{\ensuremath{-}}1613 after the soft-to-hard spectral transition}",
      journal = {\aap},
         year = 2024,
        month = jun,
       volume = {686},
          eid = {L12},
        pages = {L12},
          doi = {10.1051/0004-6361/202450566},
archivePrefix = {arXiv},
       eprint = {2404.19601},
 primaryClass = {astro-ph.HE},
       adsurl = {https://ui.adsabs.harvard.edu/abs/2024A&A...686L..12P}
}

@BOOK{Chandrasekhar1960ratr.book.....C,
       author = {{Chandrasekhar}, Subrahmanyan},
        title = "{Radiative transfer}",
         year = 1960,
       adsurl = {https://ui.adsabs.harvard.edu/abs/1960ratr.book.....C}
}

@ARTICLE{Steiner2024ApJ...969L..30S,
       author = {{Steiner}, James F. and {Nathan}, Edward and {Hu}, Kun and {Krawczynski}, Henric and {Dov{\v{c}}iak}, Michal and {Veledina}, Alexandra and {Muleri}, Fabio and {Svoboda}, Jiri and {Alabarta}, Kevin and {Parra}, Maxime and {Bhargava}, Yash and {Matt}, Giorgio and {Poutanen}, Juri and {Petrucci}, Pierre-Olivier and {Tennant}, Allyn F. and {Baglio}, M. Cristina and {Baldini}, Luca and {Barnier}, Samuel and {Bhattacharyya}, Sudip and {Bianchi}, Stefano and {Brigitte}, Maimouna and {Cabezas}, Mauricio and {Cangemi}, Floriane and {Capitanio}, Fiamma and {Casey}, Jacob and {Rodriguez Cavero}, Nicole and {Castellano}, Simone and {Cavazzuti}, Elisabetta and {Chun}, Sohee and {Churazov}, Eugene and {Costa}, Enrico and {Di Lalla}, Niccol{\`o} and {Di Marco}, Alessandro and {Egron}, Elise and {Ewing}, Melissa and {Fabiani}, Sergio and {Garc{\'\i}a}, Javier A. and {Green}, David A. and {Grinberg}, Victoria and {Hadrava}, Petr and {Ingram}, Adam and {Kaaret}, Philip and {Kislat}, Fabian and {Kitaguchi}, Takao and {Kravtsov}, Vadim and {Kub{\'a}tov{\'a}}, Brankica and {La Monaca}, Fabio and {Latronico}, Luca and {Loktev}, Vladislav and {Malacaria}, Christian and {Marin}, Fr{\'e}d{\'e}ric and {Marinucci}, Andrea and {Maryeva}, Olga and {Mastroserio}, Guglielmo and {Mizuno}, Tsunefumi and {Negro}, Michela and {Omodei}, Nicola and {Podgorn{\'y}}, Jakub and {Rankin}, John and {Ratheesh}, Ajay and {Rhodes}, Lauren and {Russell}, David M. and {{\v{S}}lechta}, Miroslav and {Soffitta}, Paolo and {Spooner}, Sean and {Suleimanov}, Valery and {Tombesi}, Francesco and {Trushkin}, Sergei A. and {Weisskopf}, Martin C. and {Zane}, Silvia and {Zdziarski}, Andrzej A. and {Zhang}, Sixuan and {Zhang}, Wenda and {Zhou}, Menglei and {Agudo}, Iv{\'a}n and {Antonelli}, Lucio A. and {Bachetti}, Matteo and {Baumgartner}, Wayne H. and {Bellazzini}, Ronaldo and {Bongiorno}, Stephen D. and {Bonino}, Raffaella and {Brez}, Alessandro and {Bucciantini}, Niccol{\`o} and {Chen}, Chien-Ting and {Ciprini}, Stefano and {De Rosa}, Alessandra and {Del Monte}, Ettore and {Di Gesu}, Laura and {Donnarumma}, Immacolata and {Doroshenko}, Victor and {Ehlert}, Steven R. and {Enoto}, Teruaki and {Evangelista}, Yuri and {Ferrazzoli}, Riccardo and {Gunji}, Shuichi and {Hayashida}, Kiyoshi and {Heyl}, Jeremy and {Iwakiri}, Wataru and {Jorstad}, Svetlana G. and {Karas}, Vladimir and {Kolodziejczak}, Jeffery J. and {Liodakis}, Ioannis and {Maldera}, Simone and {Manfreda}, Alberto and {Marscher}, Alan P. and {Marshall}, Herman L. and {Massaro}, Francesco and {Mitsuishi}, Ikuyuki and {Ng}, Chi-Yung and {O'Dell}, Stephen L. and {Oppedisano}, Chiara and {Papitto}, Alessandro and {Pavlov}, George G. and {Peirson}, Abel L. and {Perri}, Matteo and {Pesce-Rollins}, Melissa and {Pilia}, Maura and {Possenti}, Andrea and {Puccetti}, Simonetta and {Ramsey}, Brian D. and {Roberts}, Oliver J. and {Romani}, Roger W. and {Sgr{\`o}}, Carmelo and {Slane}, Patrick and {Spandre}, Gloria and {Swartz}, Douglas A. and {Tamagawa}, Toru and {Tavecchio}, Fabrizio and {Taverna}, Roberto and {Tawara}, Yuzuru and {Thomas}, Nicholas E. and {Trois}, Alessio and {Tsygankov}, Sergey S. and {Turolla}, Roberto and {Vink}, Jacco and {Wu}, Kinwah and {Xie}, Fei},
        title = "{An IXPE-led X-Ray Spectropolarimetric Campaign on the Soft State of Cygnus X-1: X-Ray Polarimetric Evidence for Strong Gravitational Lensing}",
      journal = {\apjl},
         year = 2024,
        month = jul,
       volume = {969},
       number = {2},
          eid = {L30},
        pages = {L30},
          doi = {10.3847/2041-8213/ad58e4},
archivePrefix = {arXiv},
       eprint = {2406.12014},
 primaryClass = {astro-ph.HE},
       adsurl = {https://ui.adsabs.harvard.edu/abs/2024ApJ...969L..30S}
}

@ARTICLE{Dovvciak2024Galax..12...54D,
       author = {{Dov{\v{c}}iak}, Michal and {Podgorn{\'y}}, Jakub and {Svoboda}, Ji{\v{r}}{\'\i} and {Steiner}, James F. and {Kaaret}, Philip and {Krawczynski}, Henric and {Ingram}, Adam and {Kravtsov}, Vadim and {Marra}, Lorenzo and {Muleri}, Fabio and {Garc{\'\i}a}, Javier A. and {Mastroserio}, Guglielmo and {Miku{\v{s}}incov{\'a}}, Romana and {Ratheesh}, Ajay and {Cavero}, Nicole Rodriguez},
        title = "{IXPE View of BH XRBs during the First 2.5 Years of the Mission}",
      journal = {Galaxies},
         year = 2024,
        month = sep,
       volume = {12},
       number = {5},
          eid = {54},
        pages = {54},
          doi = {10.3390/galaxies12050054},
       adsurl = {https://ui.adsabs.harvard.edu/abs/2024Galax..12...54D}
}

@ARTICLE{Zhang2019ApJ...875..148Z,
       author = {{Zhang}, Wenda and {Dov{\v{c}}iak}, Michal and {Bursa}, Michal},
        title = "{Constraining the Size of the Corona with Fully Relativistic Calculations of Spectra of Extended Coronae. I. The Monte Carlo Radiative Transfer Code}",
      journal = {\apj},
         year = 2019,
        month = apr,
       volume = {875},
       number = {2},
          eid = {148},
        pages = {148},
          doi = {10.3847/1538-4357/ab1261},
archivePrefix = {arXiv},
       eprint = {1903.09241},
 primaryClass = {astro-ph.HE},
       adsurl = {https://ui.adsabs.harvard.edu/abs/2019ApJ...875..148Z}
}

@ARTICLE{Krawczynski2022ApJ...934....4K,
       author = {{Krawczynski}, H. and {Beheshtipour}, B.},
        title = "{New Constraints on the Spin of the Black Hole Cygnus X-1 and the Physical Properties of its Accretion Disk Corona}",
      journal = {\apj},
         year = 2022,
        month = jul,
       volume = {934},
       number = {1},
          eid = {4},
        pages = {4},
          doi = {10.3847/1538-4357/ac7725},
archivePrefix = {arXiv},
       eprint = {2201.07360},
 primaryClass = {astro-ph.HE},
       adsurl = {https://ui.adsabs.harvard.edu/abs/2022ApJ...934....4K}
}

@INPROCEEDINGS{Novikov1973blho.conf..343N,
       author = {{Novikov}, I.~D. and {Thorne}, K.~S.},
        title = "{Astrophysics of black holes.}",
    booktitle = {Black Holes (Les Astres Occlus)},
         year = 1973,
       editor = {{Dewitt}, C. and {Dewitt}, B.~S.},
        month = jan,
        pages = {343-450},
       adsurl = {https://ui.adsabs.harvard.edu/abs/1973blho.conf..343N}
}

@ARTICLE{Munoz-Darias2011MNRAS.410..679M,
       author = {{Mu{\~n}oz-Darias}, T. and {Motta}, S. and {Belloni}, T.~M.},
        title = "{Fast variability as a tracer of accretion regimes in black hole transients}",
      journal = {\mnras},
         year = 2011,
        month = jan,
       volume = {410},
       number = {1},
        pages = {679-684},
          doi = {10.1111/j.1365-2966.2010.17476.x},
archivePrefix = {arXiv},
       eprint = {1008.0558},
 primaryClass = {astro-ph.HE},
       adsurl = {https://ui.adsabs.harvard.edu/abs/2011MNRAS.410..679M}
}

@ARTICLE{Heil2015MNRAS.448.3339H,
       author = {{Heil}, L.~M. and {Uttley}, P. and {Klein-Wolt}, M.},
        title = "{Power colours: simple X-ray binary variability comparison}",
      journal = {\mnras},
         year = 2015,
        month = apr,
       volume = {448},
       number = {4},
        pages = {3339-3347},
          doi = {10.1093/mnras/stv191},
archivePrefix = {arXiv},
       eprint = {1405.2024},
 primaryClass = {astro-ph.HE},
       adsurl = {https://ui.adsabs.harvard.edu/abs/2015MNRAS.448.3339H}
}

@ARTICLE{Ebisawa2003ApJ...597..780E,
       author = {{Ebisawa}, Ken and {{\.Z}ycki}, Piotr and {Kubota}, Aya and {Mizuno}, Tsunefumi and {Watarai}, Ken-ya},
        title = "{Accretion Disk Spectra of Ultraluminous X-Ray Sources in Nearby Spiral Galaxies and Galactic Superluminal Jet Sources}",
      journal = {\apj},
         year = 2003,
        month = nov,
       volume = {597},
       number = {2},
        pages = {780-797},
          doi = {10.1086/378586},
archivePrefix = {arXiv},
       eprint = {astro-ph/0307392},
 primaryClass = {astro-ph},
       adsurl = {https://ui.adsabs.harvard.edu/abs/2003ApJ...597..780E}
}

@ARTICLE{Zhou2019PhRvD..99j4031Z,
       author = {{Zhou}, Menglei and {Abdikamalov}, Askar B. and {Ayzenberg}, Dimitry and {Bambi}, Cosimo and {Liu}, Honghui and {Nampalliwar}, Sourabh},
        title = "{XSPEC model for testing the Kerr black hole hypothesis using the continuum-fitting method}",
      journal = {\prd},
         year = 2019,
        month = may,
       volume = {99},
       number = {10},
          eid = {104031},
        pages = {104031},
          doi = {10.1103/PhysRevD.99.104031},
archivePrefix = {arXiv},
       eprint = {1903.09782},
 primaryClass = {gr-qc},
       adsurl = {https://ui.adsabs.harvard.edu/abs/2019PhRvD..99j4031Z}
}

@ARTICLE{Malzac2009MNRAS.392..570M,
       author = {{Malzac}, Julien and {Belmont}, Renaud},
        title = "{The synchrotron boiler and the spectral states of black hole binaries}",
      journal = {\mnras},
         year = 2009,
        month = jan,
       volume = {392},
       number = {2},
        pages = {570-589},
          doi = {10.1111/j.1365-2966.2008.14142.x},
archivePrefix = {arXiv},
       eprint = {0810.4458},
 primaryClass = {astro-ph},
       adsurl = {https://ui.adsabs.harvard.edu/abs/2009MNRAS.392..570M}
}

@ARTICLE{Magdziarz1995MNRAS.273..837M,
       author = {{Magdziarz}, Pawel and {Zdziarski}, Andrzej A.},
        title = "{Angle-dependent Compton reflection of X-rays and gamma-rays}",
      journal = {\mnras},
         year = 1995,
        month = apr,
       volume = {273},
       number = {3},
        pages = {837-848},
          doi = {10.1093/mnras/273.3.837},
       adsurl = {https://ui.adsabs.harvard.edu/abs/1995MNRAS.273..837M}
}

\end{document}